\documentclass[twocolumn]{aastex61}
\usepackage{amsmath}

\newcommand{\teq}{$T_{\rm eq}$}

\newcommand{\gerg}{Gerg s$^{-1}$ cm$^{-2}$}

\begin{document}
\title{Bayesian Analysis of Hot Jupiter Radius Anomalies: Evidence for Ohmic Dissipation?}
\author{Daniel P. Thorngren}
\affil{Department of Physics, University of California, Santa Cruz}
\author{Jonathan J. Fortney}
\affil{Department of Astronomy and Astrophysics, University of California, Santa Cruz}
\begin{abstract}
The cause of hot Jupiter radius inflation, where giant planets with \teq$>1000\;\mathrm{K}$ are significantly larger than expected, is an open question and the subject of many proposed explanations.  Many of these hypotheses postulate an additional anomalous power which heats planets' convective interiors, leading to larger radii.  Rather than examine these proposed models individually, we determine what anomalous powers are needed to explain the observed population's radii, and consider which models are most consistent with this.  We examine 281 giant planets with well-determined masses and radii and apply thermal evolution and Bayesian statistical models to infer the anomalous power as a fraction of (and varying with) incident flux $\epsilon(F)$ that best reproduces the observed radii.  First, we observe that the inflation of planets below about $M=0.5\;\mathrm{M}_\mathrm{J}$ appears very different than their higher mass counterparts, perhaps as the result of mass loss or an inefficient heating mechanism.  As such, we exclude planets below this threshold.  Next, we show with strong significance that $\epsilon(F)$ increases with \teq towards a maximum of $\sim2.5\%$ at $T_{\rm{eq}}\approx1500\;\mathrm{K}$, and then decreases as temperatures increase further, falling to $\sim0.2\%$ at $T_\mathrm{eff}=2500$ K.  This high-flux decrease in inflation efficiency was predicted by the Ohmic dissipation model of giant planet inflation but not other models.  We also show that the thermal tides model predicts far more variance in radii than is observed.  Thus, our results provide evidence for the Ohmic dissipation model and a functional form for $\epsilon(F)$ that any future theories of hot Jupiter radii can be tested against.
\end{abstract}

\section{Introduction}\label{introduction}
The longest standing open question in exoplanetary physics is what causes the inflated radii of ``hot Jupiters", gas giant planets on short period orbits heated to equilibrium temperatures $T_{\rm{eq}} > 1000 \;\rm{K}$ \citep{Miller2011}.  Since the first detection of planet HD 209458b \citep{Charbonneau2000,Henry2000}, the radii of the vast majority of these transiting gas giants have exceeded the expected radius of $\sim$1.1 times that of Jupiter, sometimes approaching 2 Jupiter radii.  This excess radius appears to correlate with the level of incident stellar irradiation \citep{Guillot2002,Laughlin2011}, rather than e.g. semi-major axis \citep{Weiss2013}.  A wide range of theories have been proposed to explain this, most of which postulate an additional ``anomalous" power which heats the convective interior of the planet, leading to larger radii.  Typically, these theories are tested by directly modeling the physics to determine if they can produce large enough radii to explain the observations (e.g. \cite{Tremblin2017,Ginzburg2016}).  We shall take the a more complete approach by determining what anomalous powers are needed to explain the radii of the whole observed population, and then considering what models are most consistent with this.

\begin{figure}[b!]
    \centering
    \includegraphics[width=.47\textwidth]{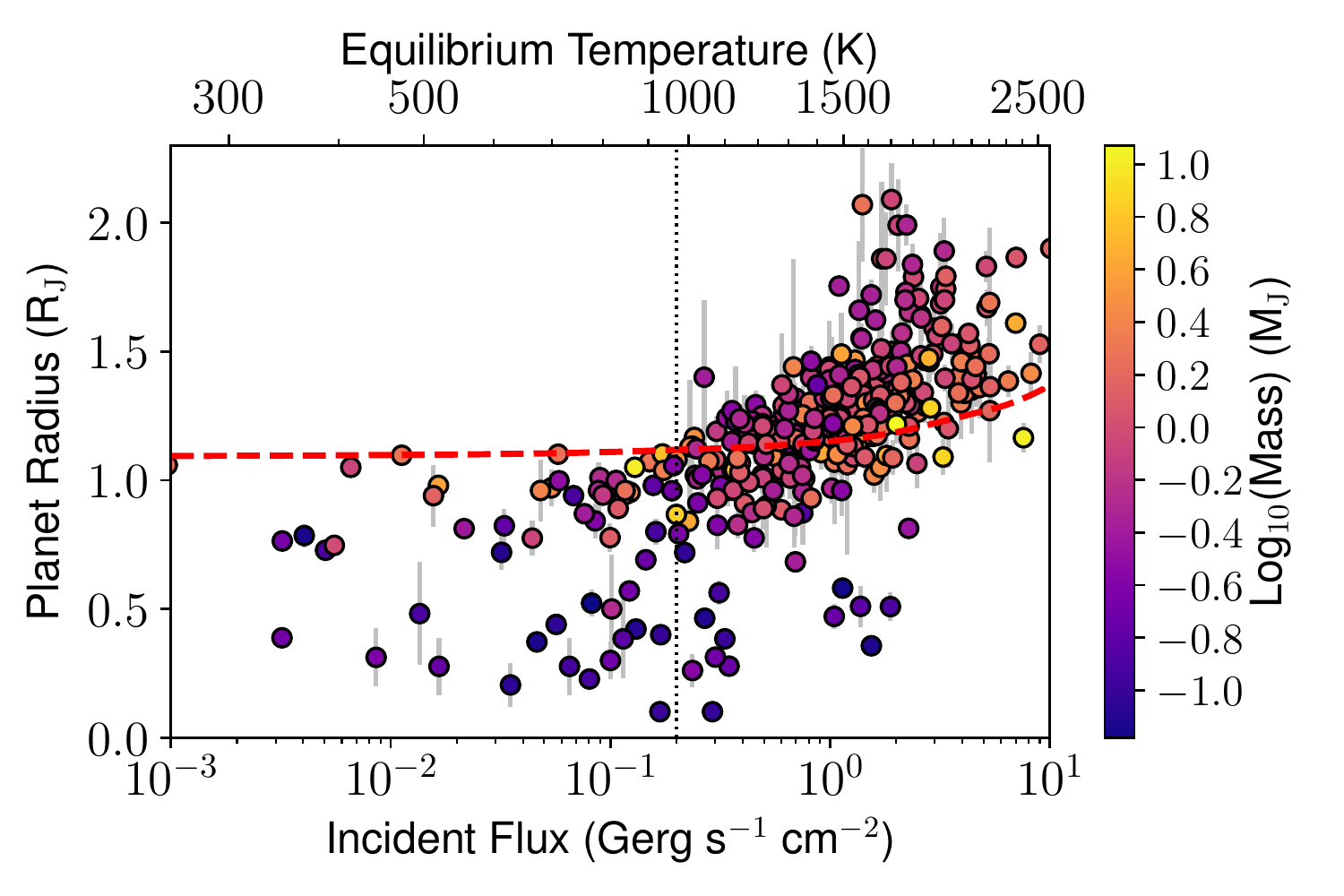}
    \caption{The radii of transiting giant exoplanets plotted against their incident flux (or equilibrium temperature) and colored by mass on the log scale.  The dotted red line is the radius of a Jupiter-mass pure H/He model with no inflation effect, an approximate upper limit on the non-inflated case.  The dotted vertical line is the empirical flux cutoff for inflation \citep{Miller2011,Demory2011}.  Beyond this level planets are anomalously large, with the excess radius correlated with flux.  Less massive planets exhibit the strongest effect.}
    \label{fig:fluxRadius}
\end{figure}

\begin{figure}[t!]
    \centering
    \includegraphics[width=.47\textwidth]{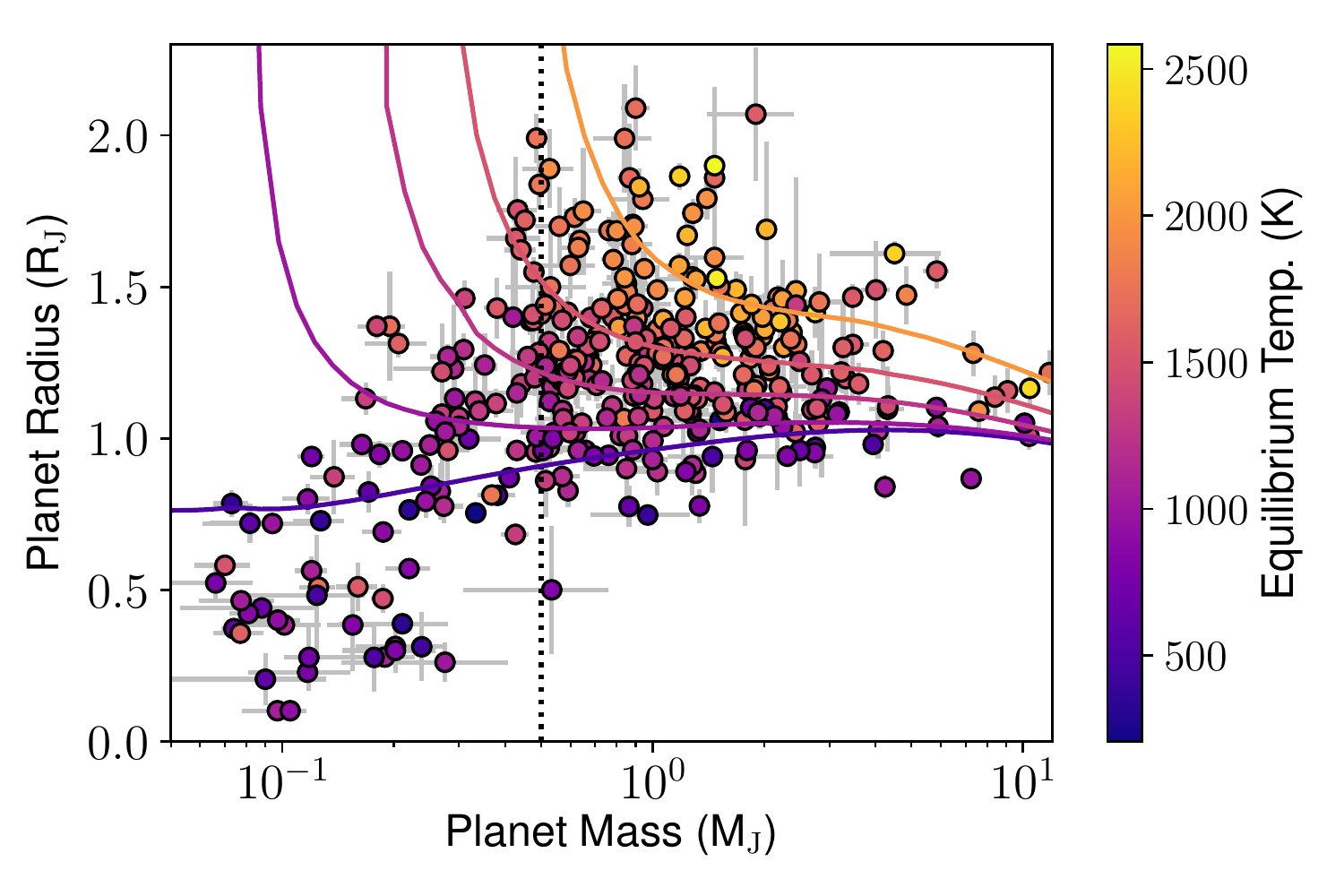}
    \caption{The radii of transiting giant exoplanets plotted against their masses, colored by equilibrium temperature.  The solid lines are the radii of model planets of average (posterior mean) composition and inflation power using our Gaussian Process results described below for various equilibrium temperatures (500, 1000, 1250, 1500, 2000 K) on the same color scale.  For each given $T_\mathrm{eq}$, models show the radii increasing dramatically at lower masses, coinciding with the absence of planets in that region.  This upturn is a feature of any plausible model of anomalous power.  Since it seems plausible that a mass-loss process affects this low-mass population, we restrict our study to planets with $M>0.5 M_J$}
    \label{fig:massRadius}
\end{figure}

This approach is feasible thanks to the work of surveys such as WASP, HAT, and Kepler, which have identified a large number of transiting giant planets.  Follow-up radial-velocity measurements have yielded mass measurements for many of these.  Merging data from the NASA Explanet Archive \citep{Akeson2013} and exoplanet.eu \citep{Schneider2011}, we examine the set of transiting planets with measured masses and radii with relative uncertainties of less than 50\%, in the mass range  $20\;\mathrm{M_\oplus} < M < 13\;\mathrm{M_J}$.  The resulting flux-radius-mass data is shown in Figure \ref{fig:fluxRadius}.  Several patterns are apparent.  First, many planets with high incident flux are anomalously large -- these are the hot Jupiters.  The flux at which this the excess radii become apparent has been estimated to occur at 0.2 \gerg \citep{Miller2011}, equivalent to an equilibrium temperature $T_{\rm{eq}} \approx 1000 $ K.  Second, the degree of radius inflation increases steadily with flux.  Finally, the degree of radius inflation is greater at lower masses.  This is more visible in Figure \ref{fig:massRadius}, which plots radius against planetary mass.

In modeling the interior structure of a transiting giant planet with a measured mass, there are two key variables which are not directly observable: the bulk heavy-element abundance and the anomalous power. Planets at fluxes below the inflation threshold, including Jupiter and Saturn, are well described by evolution models with zero anomalous power.  In this cool giant regime, we can directly infer the heavy-element mass from the observables.  Our previous work, \cite{Thorngren2016}, did this for the $\sim$50 known cool transiting giant planets (those with $T_{\rm{eq}} < 1000 $ K), and observed a correlation between the planetary heavy-element mass and the total planet mass of $(M_\mathrm{z}/\mathrm{M_\oplus}) \approx 58 (M/\mathrm{M_J})^{.61}$.  That cool giant sample and this hot giant sample do not differ much in semi-major axis (typically $\sim.1$ vs. $\sim.03$ AU), so we do not expect that their formation mechanisms or composition trends to differ.  Thus for this work, we apply this relation with its predictive uncertainty as a population-level prior on the heavy-element masses of the hot Jupiters.  By doing this, we constrain one of the two unobserved variables, allowing us to infer planetary anomalous power.  Individually, planets may vary in composition so by themselves our predictions are not particularly precise.  However, since planets as a population will follow the trend line, a hierarchical Bayesian model based on this prior allows us to combine information from our whole sample to infer the shape of the anomalous power as a function of the flux $\epsilon(F)$.  The use of the flux as a predictor was suggested by \cite{Guillot2002} and \cite{Weiss2013}, among others.

A key advantage of this approach is that it is robust against certain sources of modeling error.  In \cite{Thorngren2016}, we discussed the modest systematic uncertainties inherited from the equations of state and the distribution of metals within the planet (e.g. core vs mixed into the envelope).  These issues, as well as statistical uncertainty regarding the mass-metallicity trend and our use of fixed-metallicity atmospheres, could lead to an error in the radius of the model planets.  Two factors would act to ameliorate these effects.  First, the effects of radius suppression from metallicity would act on planets regardless of temperature, and so the first order errors in deriving the mass-metallicity trend and the impact of metals on hot giant radii would cancel out.  Second, because our sample contains a broad cross-section of different masses and fluxes for $M>0.5M_J$, biases which relate to the planet mass such as atmospheric metallicity are evenly applied to all flux levels.  Thus, this type of error may impact the overall magnitude of $\epsilon(F)$, but will have much less effect on the shape of the function.  These features do not eliminate systematic error, but they do allow for more confidence in our results.

\section{Lack of Inflated Sub-Saturns}\label{subSaturns}
\begin{figure}
    \centering
        \includegraphics[width=.47\textwidth]{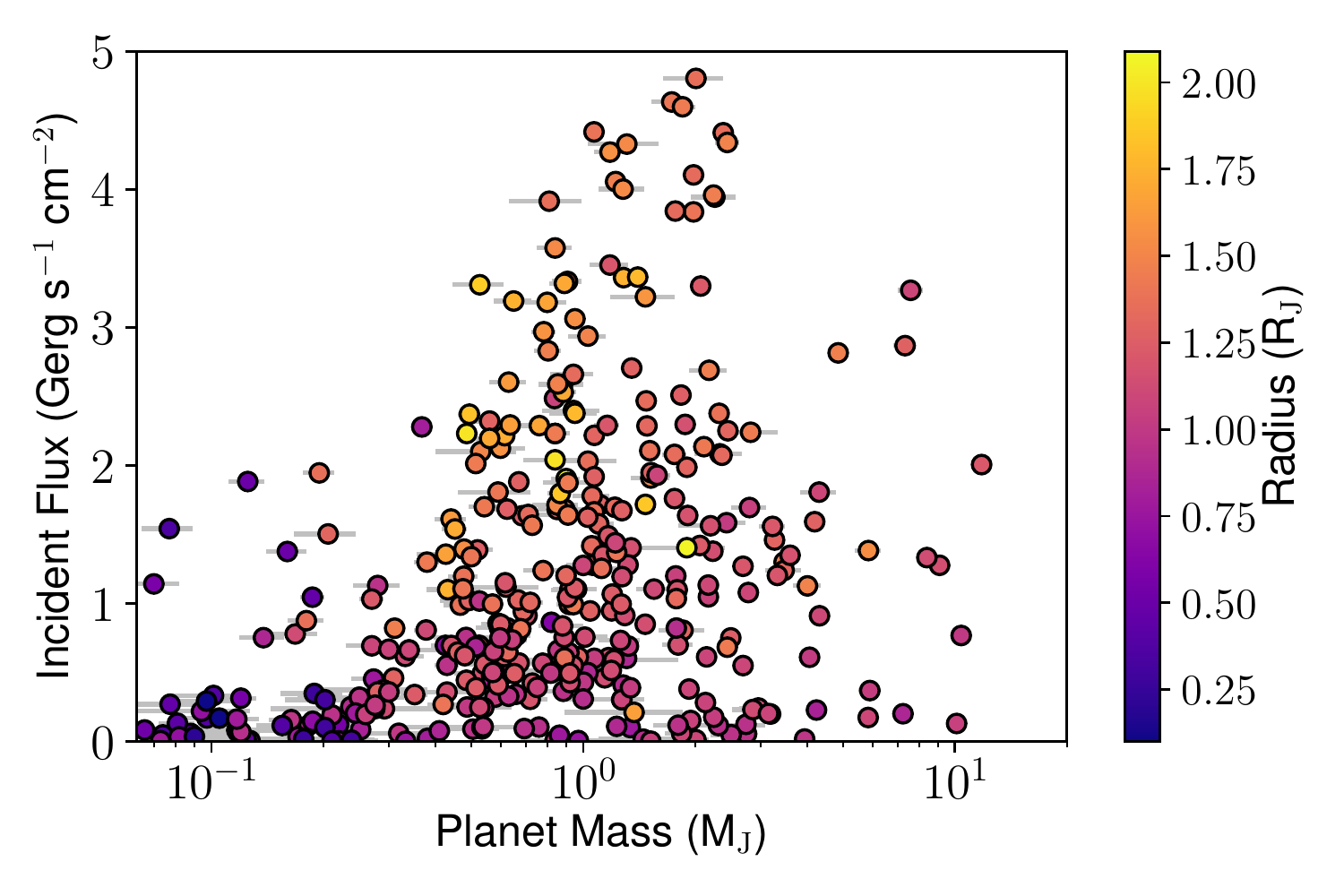}
    \caption{The mass vs.~flux of observed transiting giant planets, colored by radius.  Below about 0.4 $M_J$, considerably fewer high-flux planets are detected, an effect not seen in low-flux planets.  Transit observational biases do not explain this.  Runaway mass loss could explain both this and the lack of low-mass highly-inflated planets, though biases from formation and migration models might also exist.}
    \label{fig:massflux}
\end{figure}

An interesting feature is apparent in the mass radius relationship.  Figure \ref{fig:massRadius} shows the masses and radii of our sample of planets, along with prediction lines of constant temperature and inflation power.  The relationship between the temperature (color) and inflation power is posterior to our model (discussed later), but the general shape of the lines themselves is generic, and appears for any mass-independent model of inflation power.  It is apparent that with decreasing mass and constant inflation power, the radius anomaly becomes larger exponentially.  This is not seen in the observed planet radii.  In fact, giant planets are not observed with surface gravity less than about 3 m/s$^2$, even though our models allow it and the transits of such large planets would be readily detectable.  This might be the result of an inflation mechanism which is inefficient at low masses, but this possibility is weakened by examining the frequency of planets in mass-flux space (see Fig. \ref{fig:massflux}).

Consider the population of high-mass Jupiters compared to lower-mass Saturns, separating the groups at $0.5 \mathrm{M_J}$.  Among Jupiters, many high-flux planets are observed: 58\% (164/281) have more than 1 \gerg.  Among Saturns, we find only 22\% (21/97) which experience this level of insolation.  This discrepancy does not appear to result from any observational biases.  It is possible that significant mass loss could occur if planets inflate too much.  Because radii increase with decreasing mass, any mass loss that occurs might experience positive feedback.  This  is similar to what was seen in \cite{Baraffe2004}, though their mass loss rate appears to have been too high \citep{Hubbard2007}.  The best alternative hypothesis appears to be that Saturns preferentially stop migration further from the parent star and that planets at these masses also experience a significantly less efficient inflation effect.  Further study will require more advanced models, which we leave to future work.  To avoid this issue, we restrict our attention to planets with $M > 0.5 M_\textrm{J}$.

\section{Planet Models}\label{planetModels}
Our interior structure models are broadly the same as those in \cite{Thorngren2016}, with only two changes for this work on inflated giant planets.  We solve the equations of hydrostatic equilibrium, conservation of mass, and an equation of state (EOS) based on the SCvH (\cite{Saumon1995}) solar H/He EOS and the EOS of a 50/50 ice/rock mixture \citep{Thompson1990}.

\begin{gather}
\frac{\partial P}{\partial m} = -\frac{G m}{4 \pi r^4}\\
\frac{\partial r}{\partial m} = \frac{1}{4 \pi r^2 \rho}\\
\rho = \rho(P,T)
\end{gather}

\begin{figure}
    \centering
    \includegraphics[width=.47\textwidth]{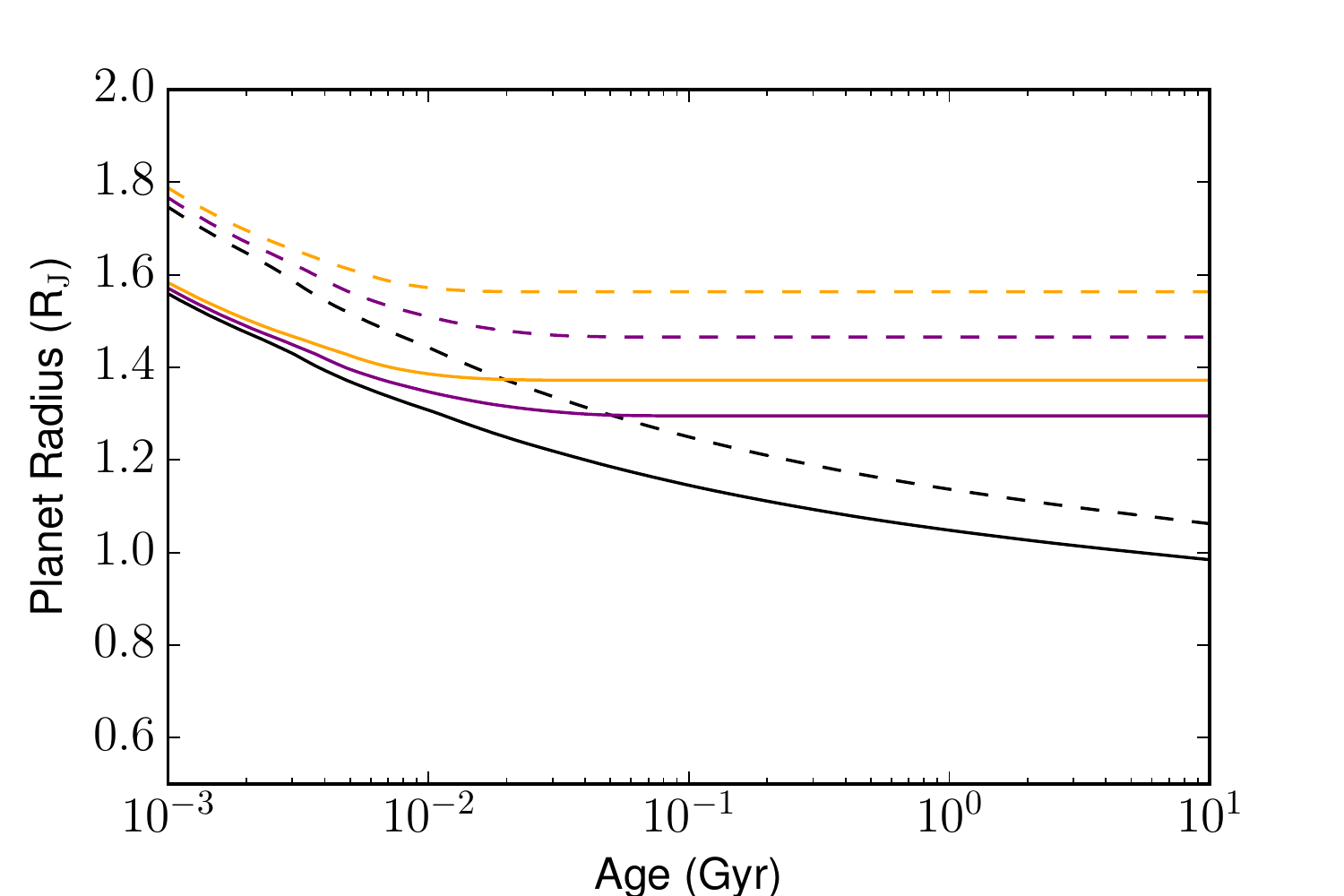}
    \caption{Example outputs of our evolution models for a 1 $M_J$ planet at 2 \gerg for different heavy-element masses and values of heating efficiency.  Solid and dashed lines have 60 and 30 $M_\oplus$ of heavy elements respectively, and black, purple, and orange lines have 0, 1, and 2\% heating efficiencies respectively.  The plot extends to extremely young ages to illustrate the transition from rapidly cooling young planets to the nearly static older planets.  Planets in our sample are generally older than a gigayear, so the effects of the heavy-element abundance and heating efficiency are not easily disentangled.}
    \label{fig:evolution}
\end{figure}

Metals were fully mixed into the convective envelope using the additive volumes approximation.  No core was included because for planets of this mass the radius difference would be minor (see \cite{Thorngren2016}).  Heat flow out of the planet (and therefore thermal and structural evolution) was regulated using the atmospheric models of \cite{Fortney2007}.  Additional details and analysis of the effect of our modeling choices can be found in \cite{Thorngren2016}.  Sample evolution calculations are shown in Figure \ref{fig:evolution}.

The most important modeling addition is the inclusion of an additional heating power $\epsilon F\pi R^2$.  The resulting power balance of the interior of the planet is
\begin{equation}
    \frac{\partial E}{\partial t} = \pi R^2 (\epsilon F - 4 F_{int})
\end{equation}
Here $F_{int}$ is the intrinsic flux of energy radiated out of the planet as computed by the atmosphere model.  Note that our definition of $\epsilon$ differs slightly from other authors, such as \cite{Komacek2017}, who deposit the energy at a particular depth within the planet.  Using their results, our definitions agree for their models where the power is deposited at the radiative-convective boundary or deeper.  Otherwise, our $\epsilon$ is smaller than theirs by a factor $<1$ depending on depth and stage of evolution.

The other change was an improvement to the thermal evolution integration system.  The new system uses the SciPy \citep{Scipy} function Odeint to adaptively integrate the changes in planet internal entropy.  We have also added a system to detect when the planet is near thermal equilibrium (when $ \epsilon F \pi R^2 \approx L_{int}$), and quickly completes the evolution accordingly.  This serves to handle the stiffness of the ODE near an equilibrium of high specific entropy.

\section{Bayesian Statistical Analysis}\label{Bayesian Statistical Analysis}
Our statistical analysis is based on a hierarchical Bayesian approach, with two levels in the hierarchy.  The lower level consists of our beliefs about the properties of individual planets given the observations and our planetary mass-metallicity relation from \cite{Thorngren2016} as a prior on bulk metallicity.  The upper level combines information about the individual planets to infer population level patterns in anomalous power.  The variables we will use are listed and described in Table \ref{varsTable}.

\begin{table*}
\centering
\begin{tabular}{|c|l|}
    \hline
    \multicolumn{2}{|c|}{Parameters} \\
    \hline
    $M_z^i$, $\vec{M}_z$ &
        The bulk heavy-element mass of the $i^\textrm{th}$ planet, all planets. \\
    $M^i$, $\vec{M}$ &
        The true mass of the $i^\textrm{th}$ planet, all planets. \\
    $t^i$, $\vec{t}$ &
        The true age of the $i^\textrm{th}$ planet, all planets. \\
    $\epsilon^i$, $\vec{\epsilon}$ & 
        The anomalous heating efficiency (see section \ref{planetModels}) of the $i^\textrm{th}$ planet, all planets. \\
    \hline \hline
    \multicolumn{2}{|c|}{Hyperparameters} \\
    \hline
    $\vec{\phi}_p = [\epsilon_0, k]$ &
        The vector of hyperparameters for the power-law model of $\epsilon(F)$.\\
    $\vec{\phi}_l = [\epsilon_0, F_0, k] $
        & The vector of hyperparameters for the logistic function model of $\epsilon(F)$.\\
    $\vec{\phi}_g = [\epsilon_0, F_0, s]$ &
        The vector of hyperparameters for the Gaussian model of $\epsilon(F)$.\\
    $\vec{\phi}_{gp} = [\sigma_1,l]$ &
        The vector of hyperparameters for the Gaussian process model of $\epsilon(F)$.\\
    \hline \hline
    \multicolumn{2}{|c|}{Constants} \\
    \hline
    $\alpha$, $\beta$, $\sigma_z$ &
        Fitted values from the planetary mass-metallicity relationship.\\
    $R_{obs}^i$, $\sigma_{r}^i$ &
        The observed radius and uncertainty of the $i^\textrm{th}$ planet.\\
    $M_{obs}^i$, $\sigma_m^i$ & 
        The observed mass and uncertainty of the $i^\textrm{th}$ planet.\\
    $t_0^i$, $t_1^i$ & 
        The observational lower and upper limits on the age of the $i^\textrm{th}$ planet. \\
    $F^i$ &
        The time-average incident flux onto the $i^\textrm{th}$ planet.\\
    \hline
\end{tabular}
\caption{A list of variables used in the Bayesian model.  The superscript is the index of the planet (numbered 1 to $N=281$), whereas an arrow refers to the variable for all of the planets as a vector.  E.g. the $10^\mathrm{th}$ component of $\vec{M}_z$ is $M_z^{10}$.  Parameters refers to model parameters of the lower hierarchical level of the model, and hyperparameters refers to those of the upper level.  Constants are known, fixed values which describe the results of previous studies, and so do not need to be sampled.  $\alpha$, $\beta$, and $\sigma_z$ are from \cite{Thorngren2016} and the remainder are from various telescope-based observational studies retrieved from on exoplanets.eu \citep{Schneider2011} and the NASA Exoplanet archive \citep{Akeson2013} (see Section \ref{introduction}).}
\label{varsTable}
\end{table*}

\subsection{Planetary Statistical Models}\label{S:statsModels}
We wish to understand the observed radii of giant planets, which have normally distributed errors, in terms of our interior structure models $R(t,M_z,M,\epsilon,F)$.  As such, we construct the following normal likelihood for observing the $i^\textrm{th}$ planet's radius to be $R_{obs}$ given the structure models parameters:
\begin{multline}
    p\left(
        R_{obs}^{i}\middle|t^{i},M^{i}_z,M^{i},\epsilon^{i}
    \right) =\\
    \mathcal{N}\left(
        R_{obs}^{i}\middle|R(t^{i},M_z^{i},M^{i},\epsilon^{i},F^{i}),\sigma_r^{i}
    \right)
\end{multline}

Here $\mathcal{N}$ refers to the normal distribution, and $\mathcal{N}(x|\mu,\sigma)$ is the PDF of $\mathcal{N}(\mu,\sigma)$ evaluated at $x$ (similarly for the uniform $\mathcal{U}$ and log-normal $\mathcal{LN}$ distributions). The observed flux $F^i$ is known to a sufficient accuracy (compared to the other observations) that we will neglect the effect its uncertainty has on the model radius uncertainty.  Previous studies provide us with observational constraints on $M_i$ and $t^i$, which we will use as priors.  Combined with the motivated prior on $M_z^i$ from our mass metallicity relationship, we have:

\begin{gather}
    t^i \sim \mathcal{U}(t_0^i,t_1^i)\\
    M_z^i \sim \mathcal{LN}\left(
        \alpha + \beta \log(M^i),\sigma_z
    \right)\\
    M^i \sim \mathcal{N}\left(
        M_{obs}^i,\sigma_m^i
    \right) \label{mzPrior}
\end{gather}

 $\mathcal{LN}(\mu,\sigma)$ is the base-10 log-normal distribution with location $\mu$ and scale $\sigma$ (i.e. the $\log_{10}$ of the variable is distributed as $\mathcal{N}(\mu,\sigma)$).  Using these priors, we can write a posterior distribution for the structure model parameters ($t^i$, $M_z^i$,$M^i$, $\epsilon^i$) as follows:
\begin{align}
    p(&t^i,M_z^i,M^i,\epsilon^i|R_{obs}^i) \nonumber\\
    =&p(R_{obs}^i|t^i,M_z^i,M^i,\epsilon^i)
        p(t^i,M_z^i,M^i,\epsilon^i) / p(R_{obs}^i)\\
    \propto&
        p(R_{obs^i}|t^i,M_z^i,M^i,\epsilon^i)
        p(t^i)p(M_z^i|M^i)p(M^i)p(\epsilon^i)\\
    \propto& \mathcal{N}\left(
        R_{obs}^i|R(t^i,M_z^i,M^i,\epsilon^i,F^i),\sigma_{r}^i
        \right) \times \label{SinglePlanetPosterior} \\
    &\mathcal{U}(t^i|t_0^i,t_1^i) \times
        \mathcal{LN}\left(M_z^i|\alpha + \beta \log(M^i),
        \sigma_z \right) \times \nonumber\\
    &\mathcal{N}\left(M^i|M_{obs}^i,\sigma_{m}^i\right)
        \times p(\epsilon) \nonumber
\end{align}

\begin{figure*}
    \centering
    \includegraphics[width=.9\textwidth]{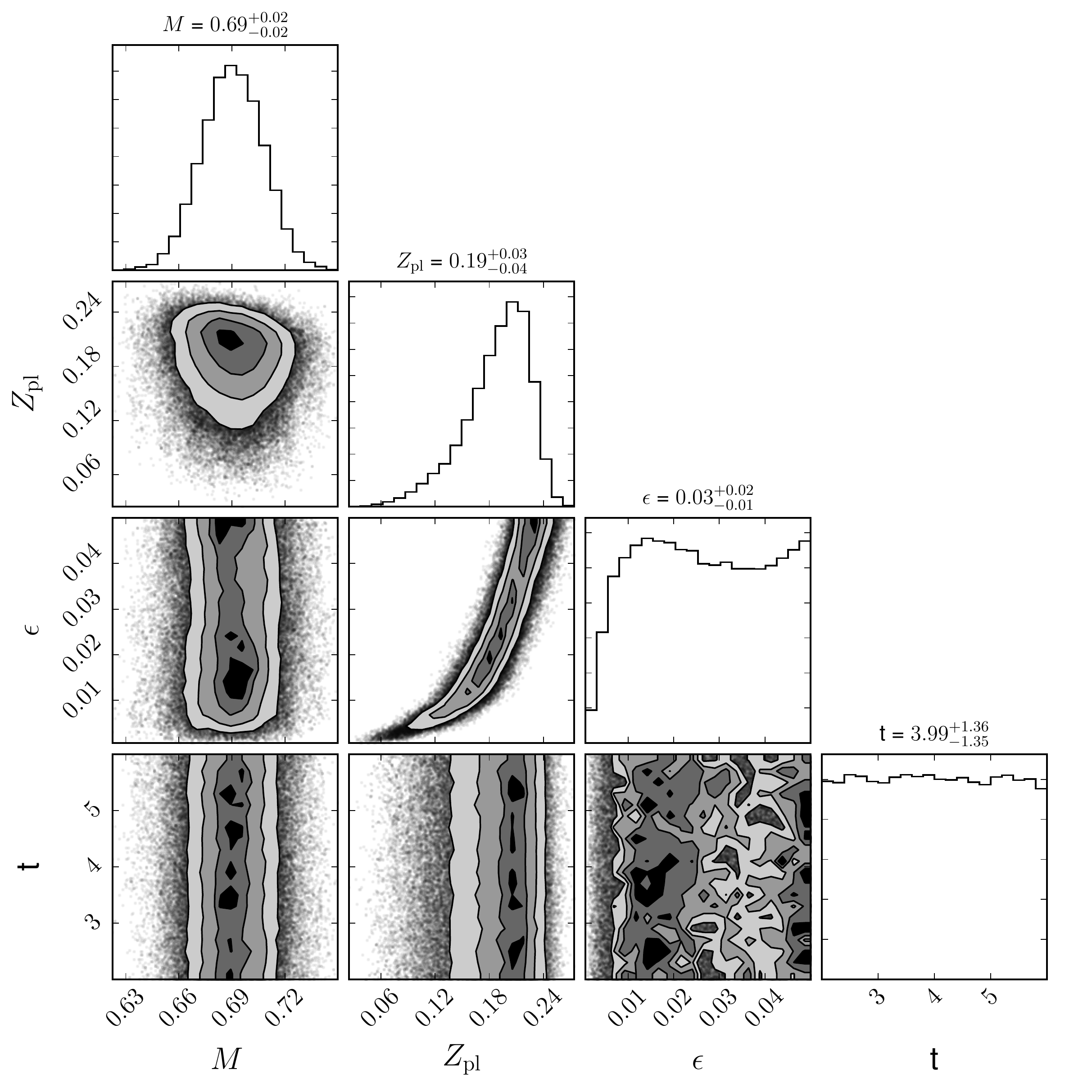}
    \caption{Inferred parameters for HD 209458 b, using equation \ref{SinglePlanetPosterior}.  The parameters are mass in Jupiter masses, planetary metal mass fraction, inflation efficiency, and age in gigayears.  The planet is old enough that its age uncertainty has little effect on the other parameters.  As expected, the main driver of $\epsilon$ uncertainty is $Z_\textrm{pl}$.  For this planet, we disfavor an inflation efficiency below $\sim1\%$.  Together with other planets, some of which disfavor high $\epsilon$, this forms the basis for our inference of $\epsilon(F)$.}
    \label{fig:planetPostExample}
\end{figure*}

The purpose of this model is to infer $\epsilon^i$.  If we apply a simple uniform prior $\epsilon^i \sim \mathcal{U}(0,5\%)$, we can infer the interior structure parameters for the $i^\mathrm{th}$ planet.  Figure \ref{fig:planetPostExample} shows the results of this approach for HD 209458 b.  Unfortunately, as seen in the figure, data from a single planet does not provide enough information to infer much about $\epsilon^i$.  In the next section, we describe a hierarchical model which combines the information from many planets to draw conclusions about the anomalous power as a function of flux $\epsilon(F)$.

\subsection{Models of Anomalous Power}
For convenience, we define the function $Q^i$ as follows:
\begin{align}
    Q^i(&t^i,M_z^i,M^i,\epsilon^i) \equiv \\
    &\mathcal{N}\left(
        R_{obs}^i|R(t^i,M_z^i,M^i,\epsilon^i,F^i),\sigma_r^i
        \right) \times \nonumber\\
    &\mathcal{U}(t^i|t_0^i,t_1^i)\times
        \mathcal{LN}\left(M_z^i|\alpha + \beta \log(M^i),
        \sigma_z\right) \times \nonumber \\
    &\mathcal{N}\left(M^i|M_{obs}^i,\sigma_m^i\right) \nonumber
\end{align}

We include the model parameters as explicit arguments, and let the constants be indicated by the index $i$.  This function reduces the right hand side of Eq. \ref{SinglePlanetPosterior} to $Q^i(t^i,M_z^i,M^i,\epsilon^i) p(\epsilon^i)$.  To combine information from many planets together, we assume that the planet parameters $t^i$, $M_z^i$, and $M^i$ as well as $R_{obs}^i$ are a priori independent between planets, and thus we can simply multiply their probabilities together.  For this equation we will leave the prior on $\epsilon^i$ in the general form $p(\vec{\epsilon})$.

\begin{align}
    p\left(\vec{t},\vec{M_z},\vec{M},\vec{\epsilon}
        \middle|\vec{R}_{obs}\right)
        \propto p(\vec{\epsilon})
        \prod_{i=1}^N Q^i(t^i,M_z^i,M^i,\epsilon^i)
        \label{allPlanetPost}
\end{align}

We can now focus on constructing models of $\epsilon^i$.  First, we consider the models in which the heating efficiency $\epsilon$ is given by a deterministic function of several hyperparameters $\vec{\phi}$.  We will refer to this function generally as $\epsilon(F^{(i)},\vec{\phi})$, and consider several specific functions (power-law, logistic, and Gaussian), differentiated by their subscripts.  These models were chosen because they all allow for low heating efficiencies at low fluxes, but exhibit differing behavior at high fluxes.  The power-law model is a classic and simple model for many astronomical phenomena, the logistic model captures the possibility that the inflation effect "turns on" at some flux, and the Gaussian model covers the case that heating efficiency declines at high flux.
\begin{align}
    \epsilon_p(F,\vec \phi_p) &= \epsilon_0 F^k\label{powerEquation}\\
    \epsilon_l(F, \vec \phi_l) &= \frac{\epsilon_0}
        {1 + (F/F_0)^{-k}} \label{logisticEquation}\\
    \epsilon_g(F, \vec \phi_g) &= \epsilon_0 \exp\left(-
        \frac{\log_{10}(F/F_0)^2}{2s^2}\right)\label{gaussEquation}
\end{align}

For each of these models, we choose the follow weakly informative proper priors for the hyperparameters:
\begin{align}
    p(\vec{\phi}_p) &\propto
        \mathcal{U}(\vec{\epsilon}|0,5\%) \times \mathcal{N}(k|0,2)\\
    p(\vec{\phi}_l) &\propto
        \mathcal{U}(\epsilon_0|0,5\%) \times \mathcal{N}(F_0|1,2) \times \mathcal{N}(k|3,1)\\
    p(\vec{\phi}_g) &\propto
        \mathcal{U}(\epsilon_0|0,5\%) \times \mathcal{LN}(F_0|1,2) \times \mathcal{LN}(s|0,2)
\end{align}

In the power-law case, the uniform distribution demands $\epsilon_0$ and $k$ be such that that no planet's $\epsilon$ leave the $[0,5\%]$ bounds.  In the logistic case, the prior on $k$ is fairly informative, demanding that the transition be somewhat similar to the scale of the data; this parameter would be poorly constrained otherwise.  Now we substitute $\epsilon(F^i,\vec{\phi})$ into Eq. \ref{allPlanetPost}, which together with the hyperpriors gives us the following posterior:
\begin{multline}
    p\left(\vec{t},\vec{M_z},\vec{M},\vec{\phi}
        \middle|\vec{R}_{obs}\right) \propto\\
    p(\vec{\phi}) \prod_{i=1}^N
        Q^i(t^i,M_z^i,M^i,\epsilon(F^i,\vec{\phi}))
\end{multline}

The Gaussian process (GP hereafter) model takes a slightly different form.  In it, we model $\log_{10}(\epsilon)$ as a GP with mean $0$ and covariance matrix $K$.  We use the squared exponential kernel with a small white noise component $\sigma_2^2 = 10^{-3}$ for numerical convenience, which amounts to a relative spread of about 7\% in linear space.  Thus, the covariance matrix for the process is given by:
\begin{gather}
    \mathbf{K}_{\mathrm{j,k}}(\vec{\phi}_{gp}) =
        \sigma_1^2 \exp\left(-\frac{\log_{10}(F_j/F_k)^2}{2l^2}\right) 
        + \sigma_2^2 \delta_{j,k}\label{gpEquation}
\end{gather}

We define some weakly informative priors for $\vec{\phi}_{gp}$ as follows:
\begin{align}
    p\left(\vec{\phi}_{gp}\right) \propto \mathcal{LN}(\sigma^2_1|0,1) \times \mathcal{LN}(l|0,1)
\end{align}

Because we do not have simple normal distributions for them, we cannot marginalize out $\vec{\epsilon}$, and instead must keep them as parameters hierarchically connected through the GP prior.  To provide an appropriate lower boundary condition on the function, we include an independent portion of the prior on $\epsilon^i$ (in combination with the GP) such that the model is:

\begin{gather}
    p(\vec{\epsilon}_{gp}) \propto \mathcal{LN}(\vec{\epsilon}|0,\mathbf{K}(\vec{F},\vec{\phi_{gp}}))
        \prod_{i=1}^N \mathcal{U}(\epsilon^i|0,5\%) G^i(\epsilon^i)\\
    G^i(\epsilon^i) \equiv \left\{
    \begin{array}{lr}
        1 & \mathrm{for }\; F_i \geq 10^8 \\
        \mathcal{LN}(\epsilon^i|-2,1) & \mathrm{for }\; F_i < 10^8 \\
    \end{array}\right.
\end{gather}

The lognormal portion is the Gaussian Process.  The $G^i$ component is useful because it sets an appropriate lower boundary condition for $\epsilon(F)$.  Experimentation reveals that this boundary condition has little effect when $F > 2$ \gerg (the region of interest); we merely include it to best represent our belief about the function for the full range of fluxes. With these priors and likelihood, Bayes Theorem yields the posterior for the GP model:
\begin{align}
    p&\left(\vec{t},\vec{M_z},\vec{M},\vec{\epsilon},\vec{\phi}_{GP}
        \middle|\vec{R}_{obs}\right) \propto \\
    &p(\vec{\phi}_{gp})\mathcal{LN}
        \left(\vec{\epsilon}\middle|\vec{0},\mathbf{K}(\vec{F},
        \vec{\phi_{gp}})\right) \times \nonumber \\
    &\prod_{i=1}^N Q^i(t^i,M_z^i,M^i,\epsilon^i)
        \mathcal{U}(\epsilon^i|0,5\%)
        G^i(\epsilon^i)\nonumber
\end{align}

Finally, we constructed a simple model for the thermal tides model of hot Jupiter inflation \citep{Arras2009}.  We adapt the scaling relations of \cite{Socrates2013}, $L \propto T_{\textrm{eq}}^3 R^4 P^{-2}$, where L is the total anomalous power, $P$ is the period, $T_{eq}$ is the equilibrium temperature, and $R$ is the planet radius.  We model this as follows, where $\epsilon_0$ is a model parameter, using the present-day radius and flux for simplicity.
\begin{align}
    \epsilon_t(F) = \epsilon_0 R^2 P^{-2} F^{-.25}
    \label{tidesModel}
\end{align}

\subsection{Statistical Computation}
We wish to use a Metropolis-Hastings MCMC \citep{Hastings1970} sampler to draw samples from the posteriors given above.  However, if we do this with no further simplifications, we will end up exploring the parameters very slowly.  This is because the models listed above have a very large number of parameters ($\sim1100$) thanks to the many nuisance parameters ($M_z^i$, $M^i$, etc) which each have one parameter per planet.  The complexity of our Metropolis-Hastings sampler scales with dimension at roughly $\mathcal{O}(d^2)$: $\mathcal{O}(d)$ posterior PDF evaluations \citep[see][]{Roberts2004} that cost $\mathcal{O}(d)$.  However, we are really only interested in $\vec{\phi}$ for the various models, plus $\vec{\epsilon}$ in the GP case.  We can save a great deal of computational effort by directly sampling marginal distribution and rewriting the posteriors as follows:
\begin{align}
    &p\left(\vec{\phi}\middle|\vec{R}_{obs}\right) \nonumber \\
    &=\int p\left(\vec{t},\vec{M_z},\vec{M},\vec{\phi}
        \middle|\vec{R}_{obs}\right)d\vec{t} d\vec{M}_z d\vec{M}\\
    &=\int p(\vec{\phi})\prod_{i=1}^N Q^i(t^i,M_z^i,M^i,
        \epsilon(F^i,\vec{\phi}))
        d\vec{t} d\vec{M}_z d\vec{M}\label{sumIntegral}\\
    &=p(\vec{\phi})\prod_{i=1}^N \int Q^i(t^i,M_z^i,M^i,\epsilon(F^i,\vec{\phi}))dt^i dM^i_z dM^i\label{integralSum}\\
    &=p(\vec{\phi})\prod_{i=1}^N \widetilde{Q}^i(\epsilon(F^i,\vec{\phi}))
\end{align}

We use $d\vec{t}$ and the like as shorthand for integration over every component of $\vec{t}$ in sequence over their full domain; Eq. \ref{sumIntegral} has 843 nested integrals!  $\widetilde{Q}^i$ is defined as $Q^i$ integrated over $t^i$, $M_z^i$, and $M^i$:
\begin{align}
    \widetilde{Q}^i(\epsilon) \equiv \int Q^i(t^i,M_z^i,M^i,\epsilon)dt^i,dM^i_z,dM^i\label{qtilde}
\end{align}

In this way, we have rewritten the $3N$ dimensional integral in Eq. \ref{sumIntegral} as $N$ separate $3$ dimensional integrals in Eq. \ref{integralSum}.  This rewrite of the posteriors is possible because the planet parameters are only connected to each other through the hierarchical prior on $\epsilon$.  The GP posterior can be simplified in a similar fashion:
\begin{align}
    p\left(\vec{\epsilon},\vec{\phi}
    \middle|\vec{R}_{obs}\right) \propto
    &p(\vec{\phi}) \mathcal{LN}\left(\vec{\epsilon}\middle|\vec{0},\mathbf{K}(\vec{F},\vec{\phi_{gp}})\right) \times \\
    &\prod_{i=1}^N \widetilde{Q}^i(\epsilon^i) \mathcal{U}(\epsilon_0|0,5\%) G(F^i,\epsilon^i) \nonumber
\end{align}

\begin{figure}
    \centering
    \includegraphics[width=.47\textwidth]{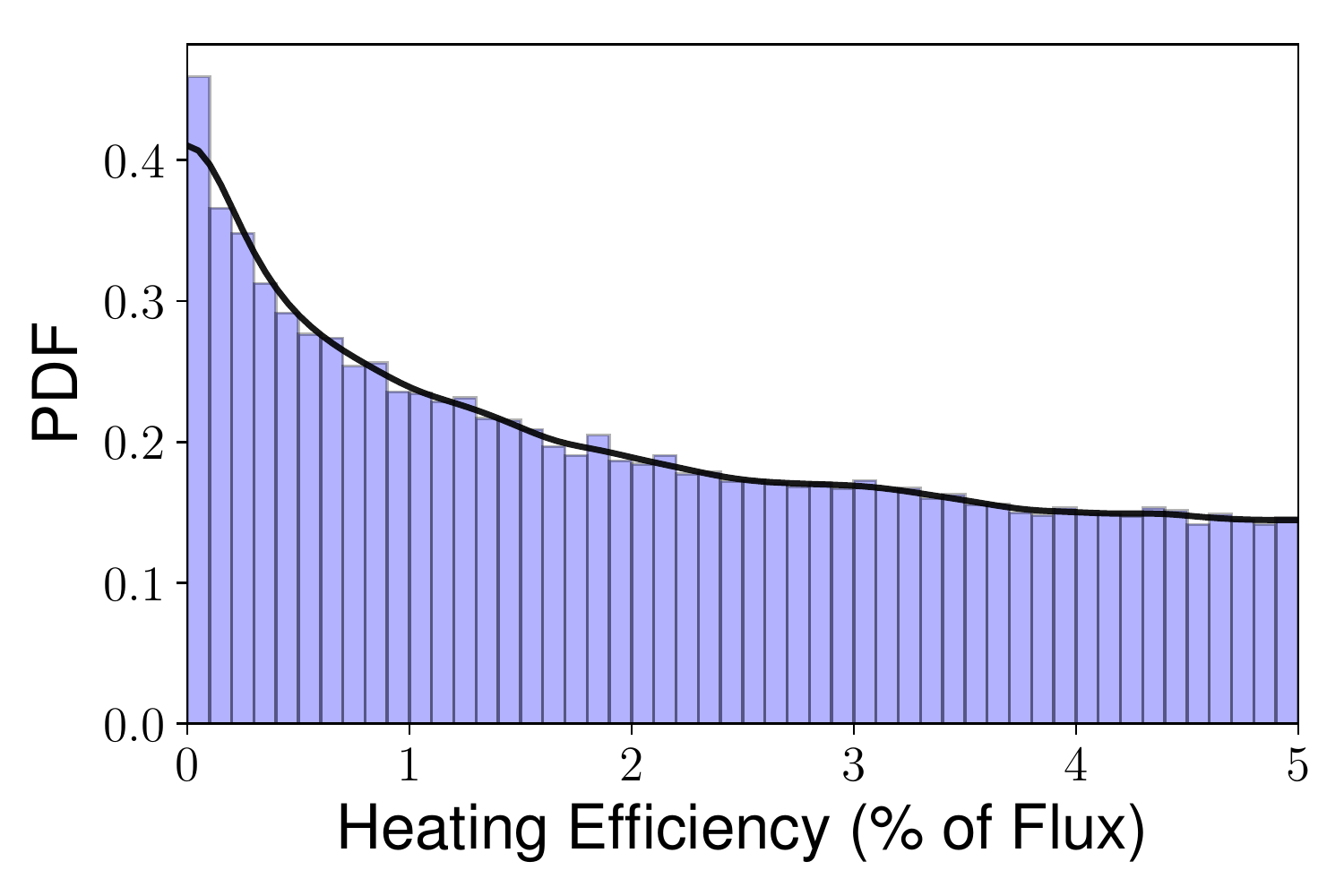}
    \caption{The histogram and kernel density estimate (the black line) of the posterior inflation power $\epsilon$ (proportional to $\widetilde{Q}^i(\epsilon)$) for WASP-43 b.  In this case, smaller values of $\epsilon$ are more likely, but larger values are not ruled out.  Note that the KDE matches the histogram, as is required for us to be able to use it as a likelihood for the upper level of the hierarchical model.}
    \label{fig:kde.pdf}
\end{figure}

Using this formulation to get posterior samples relies on our ability to compute $\widetilde{Q}^i(\epsilon)$ up to a constant of proportionality.  This is easier than it might appear.  Eq. \ref{qtilde} is proportional to the single planet posterior PDF (Eq. \ref{SinglePlanetPosterior}) for $p(\epsilon) \propto \mathcal{U}(0,5\%)$, marginalized over $t^i$, $M_z^i$, and $M^i$.  We chose this prior for epsilon because we do not believe that $\epsilon$ will exceed $5\%$.  We can estimate this marginal PDF by sampling from the posterior and applying a Gaussian kernel density estimate (KDE) with reflected boundaries \citep[see][]{Silverman1986} to the $\epsilon$ samples.  Fig. \ref{fig:kde.pdf} shows the results of this procedure forWASP-43 b.  Doing this for each planet $i$ gives us $\widetilde{Q}^i(\epsilon)$.  These can be plugged into the marginalized models (assuming $0 < \epsilon < 5\%$), radically reducing the dimension.

As estimated above, our sampler scales with dimension at roughly $\mathcal{O}(d^2)$, so breaking it up into many sub-samplers is highly desirable.  The result is a much more computationally efficient sampling system, at the cost of no longer having posterior samples of the structure parameters.

\begin{figure}[t]
    \centering
    \includegraphics[width=.47\textwidth]{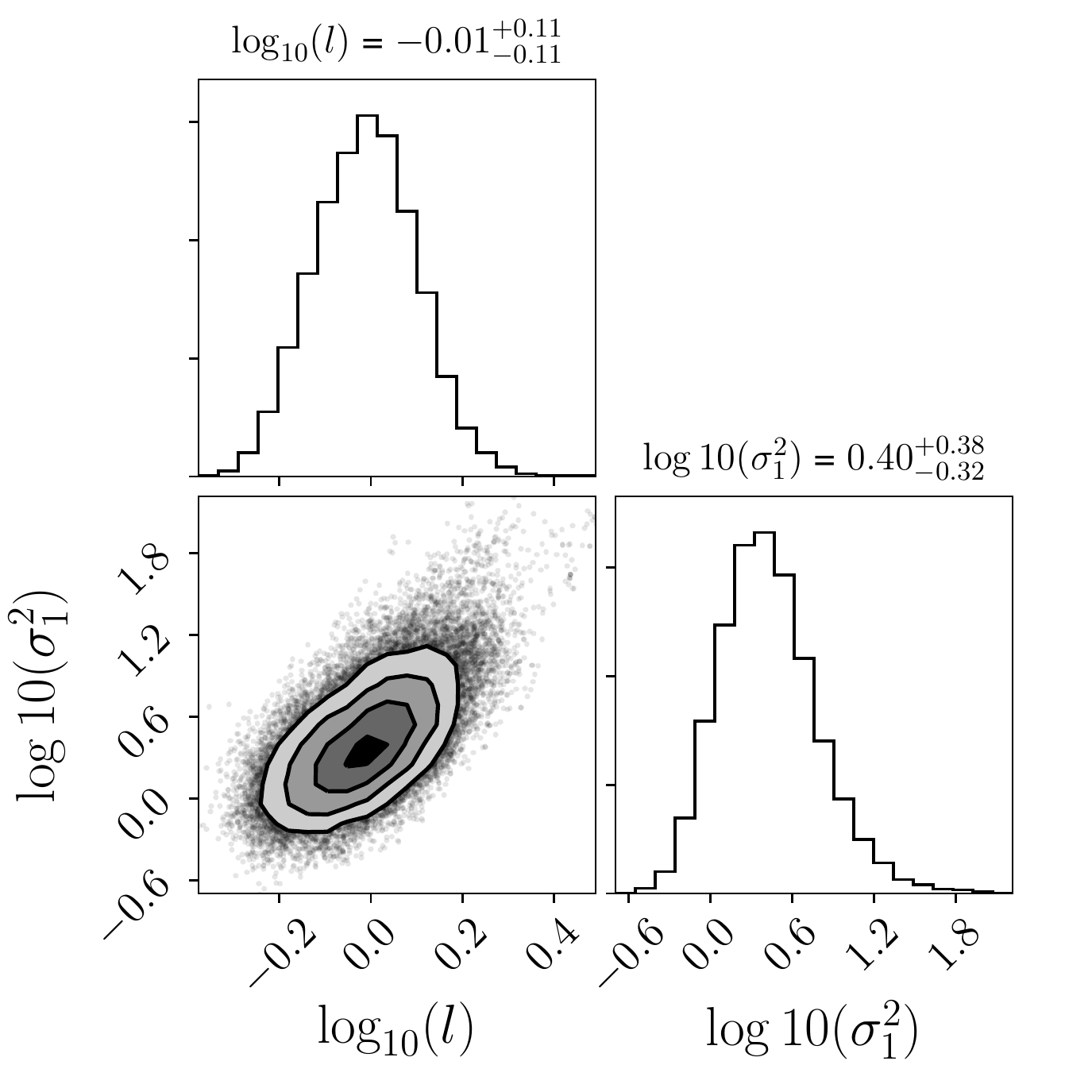}
    \caption{A scatterplot matrix of the GP hyperparameter posterior  (see eq. \ref{gpEquation}).  It is fairly well-behaved, but has a long right tail.  This is a common feature for Gaussian processes.}
    \label{fig:gpPost}
\end{figure}

\begin{figure}
    \centering
    \includegraphics[width=.47\textwidth]{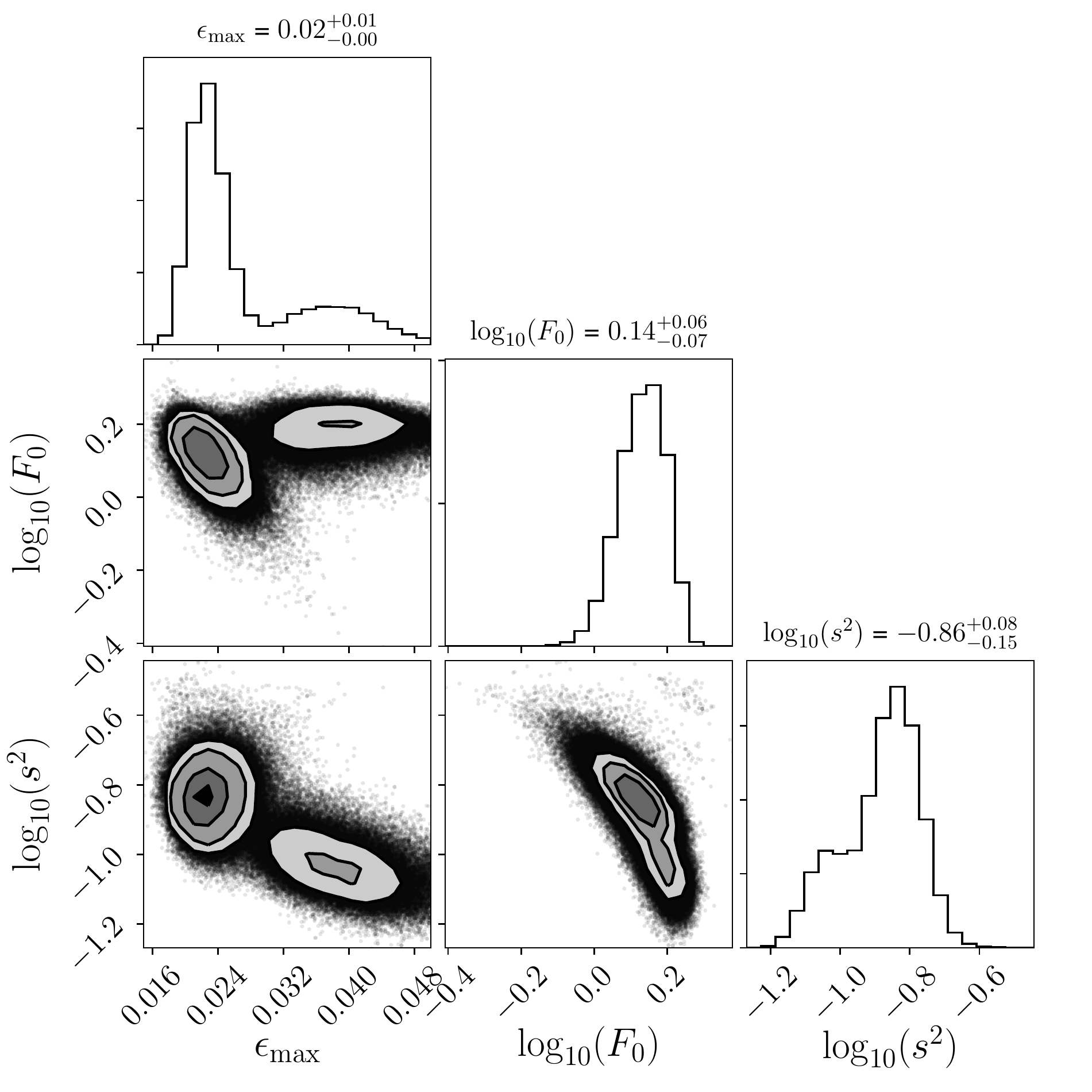}
    \caption{A scatterplot matrix of the Gaussian function hyperparameter posterior (see eq. \ref{gaussEquation}).  Two modes were observed, differing primarily in height $\epsilon_\mathrm{max}$; the model with a peak of $\epsilon\approx2\%$ is favored over the model with peak $\epsilon\approx3.5\%$ by a probability ratio of about 75\% to 25\%.  The discovery of more giant planets around the $\approx1500$ K peak will help to resolve this further.}
    \label{fig:gaussPost}
\end{figure}

\begin{figure}
    \centering
    \includegraphics[width=.47\textwidth]{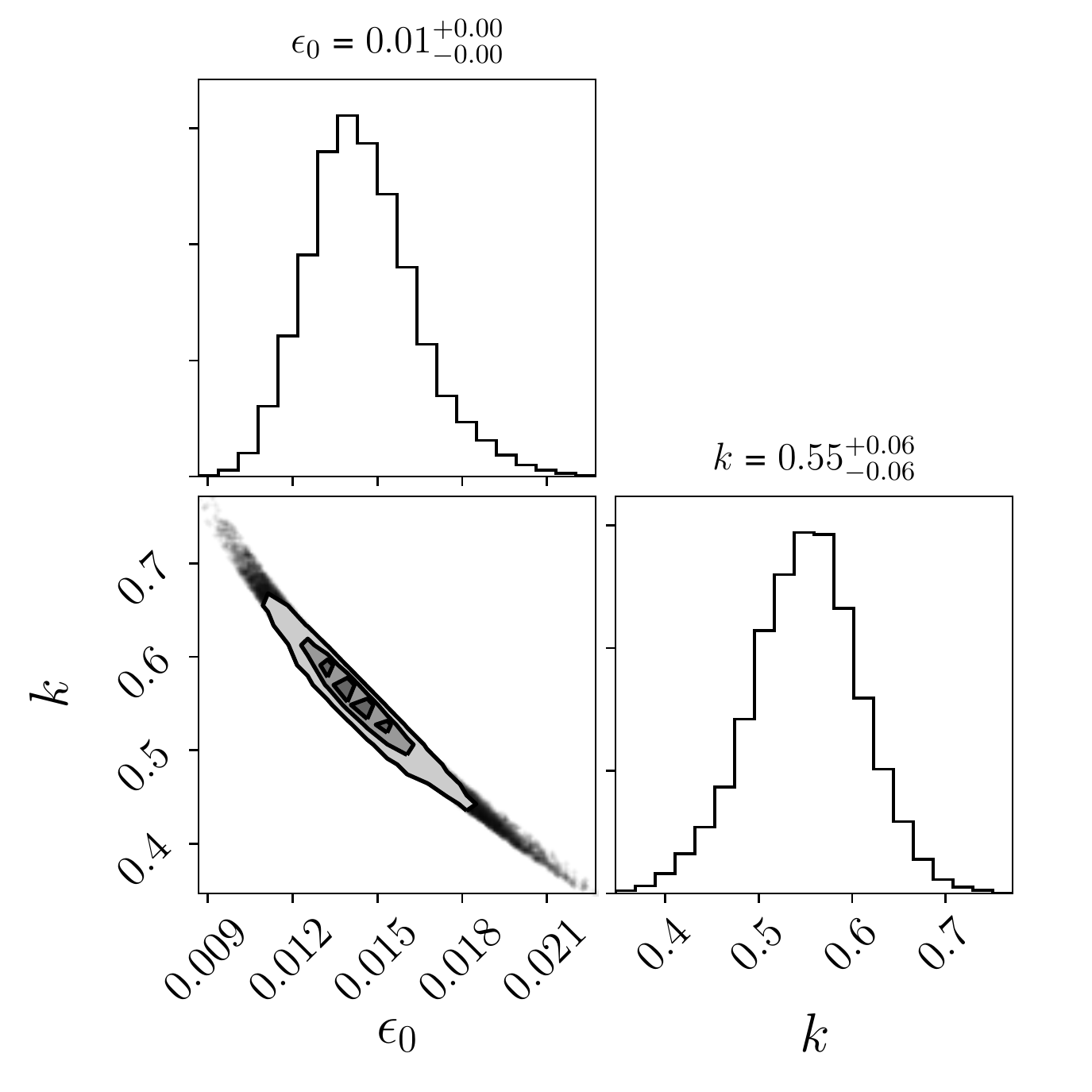}
    \caption{A scatterplot matrix of the power law hyperparameter posterior (see eq. \ref{powerEquation}.  A strong correlation between the coefficient $F_0$ and the power $k$ is seen.  This likely reflects the constraint that the function achieve adequate power for the many planets at around $T_{eq}\approx1300$ K, yet avoid exceeding 5\% for the hottest planets, which would exceed the bounds of our grid.  Such constraints are difficult for the power-law to achieve.  Regardless, as a result of its overestimate of high $T_{eq}$ radii, this model had a comparatively disfavorable DIC.}
    \label{fig:powerPost}
\end{figure}

\begin{figure}
    \centering
    \includegraphics[width=.47\textwidth]{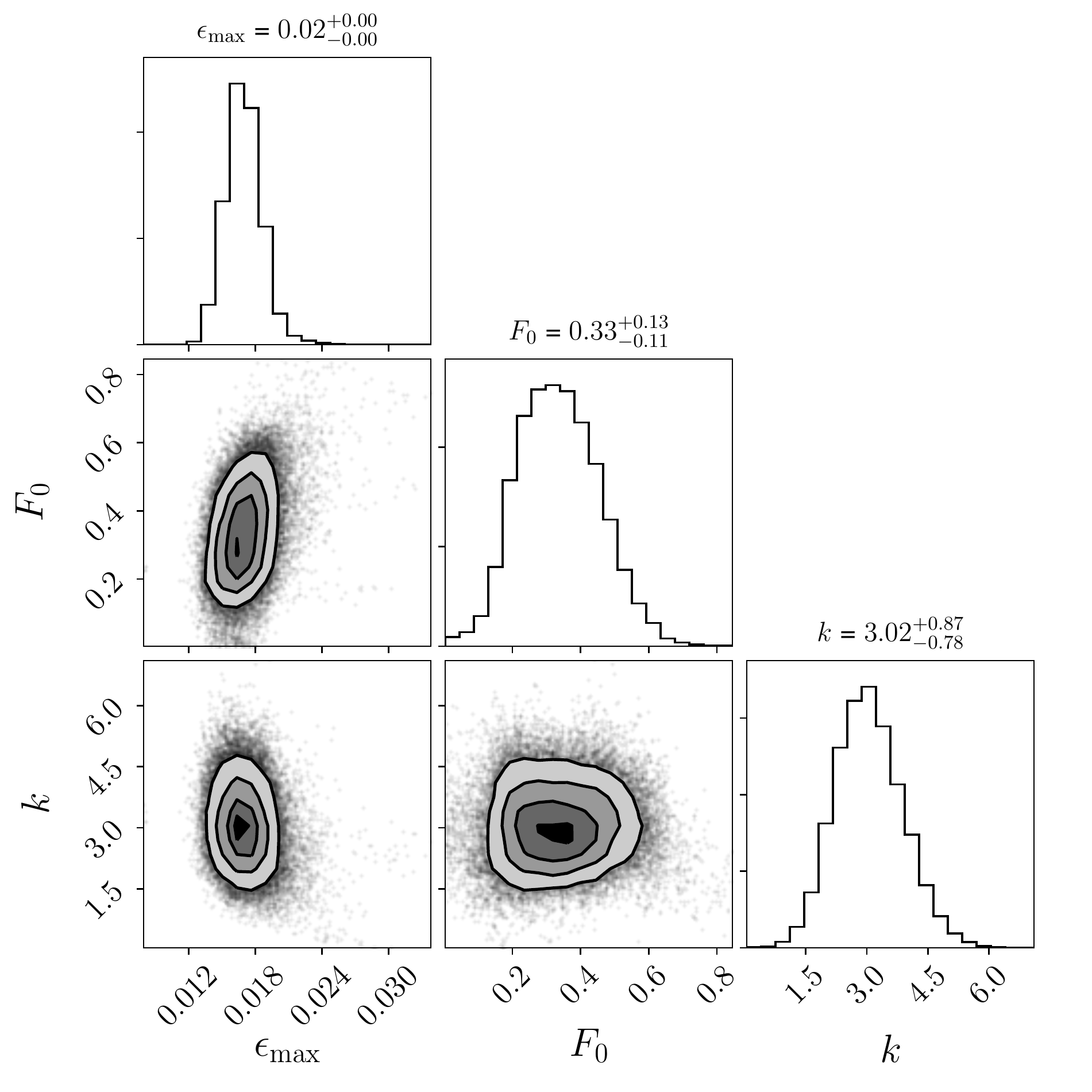}
    \caption{A scatterplot matrix of the logistic function hyperparameter posterior (see eq. \ref{logisticEquation}).  Thanks to our prior on $k$, which demanded the transition be similar to the scale of the data, the resulting posterior is well-behaved and easy to sample from.  The model is not bad, but its DIC indicates that it is still inferior to a model which decreases at high equilibrium temperatures.}
    \label{fig:logiPost}
\end{figure}

Scatterplot matrices of our upper-level model posteriors are shown in figures \ref{fig:gpPost}-\ref{fig:logiPost}, and the those of the lower level model for HD 209458 b are shown in figure \ref{fig:planetPostExample}.  The plots were made using corner.py (\cite{foreman-mackey2016}).

To compare different models, we are unable to use the more familiar model selection criteria, the BIC/AIC, as these are only defined for non-hierarchical models.  This is because in the hierarchical case the number of parameters is not well-defined \citep{Gelman2014}.  Probably the most Bayesian approach is to compare the Bayes factors (also called the evidence) of the models.  However \cite{Gelman2013} (Chapter 7.4) advise against their use in the case of continuous variables with uninformative priors as we have here.  Furthermore, computing Bayes factors here would he computationally expensive.  Instead, we make use of the Deviance Information Criterion (DIC), which is similar to the AIC in interpretation, but which makes use of an estimate of the effective number of parameters \citep{Spiegelhalter2002}, derived from the variance of the log posterior likelihood.  The empirical DIC from a set of samples is:

\begin{gather}
    \mathrm{DIC} = -2\log\left(p(\mathbf{y}|\hat{\epsilon})\right) + 4\mathrm{Var}_s\left[\log(p(\mathbf{y}|\vec{\epsilon}_s))\right]
\end{gather}

Here, $\hat{\epsilon}$ is the posterior mean of $\vec{\epsilon}$ and Var$_s$ is the variance of the log likelihood across samples.  Note that while the samples in question are taken using the posterior, this computation is done using the likelihood.  In the results, the model with the more negative DIC is favored.  The interpretation of $\Delta\mathrm{DIC}$ is similar to that of the AIC and BIC, in which differences of $> \sim 6$ are strong evidence in favor of the model with the lesser DIC (e.g. \cite{Kass1995} for BIC).

To produce posterior predictive mass-flux-radius relations, we assume the planets are old (5 Gyr.), and for given $M$ and $F$, we draw $M_z$ from Eq. \ref{mzPrior} and $\epsilon_i$ from $\epsilon(F,\vec\phi)$ marginalized over the posterior $p(\vec\phi|\vec{R}_{obs})$.  These sampled values are then plugged into the structure models $R(t,M_z,M,\epsilon,F)$.  The result is a probability distribution in $R$ for the given parameters.

\section{Results}
\begin{figure}
    \centering
    \includegraphics[width=.47\textwidth]{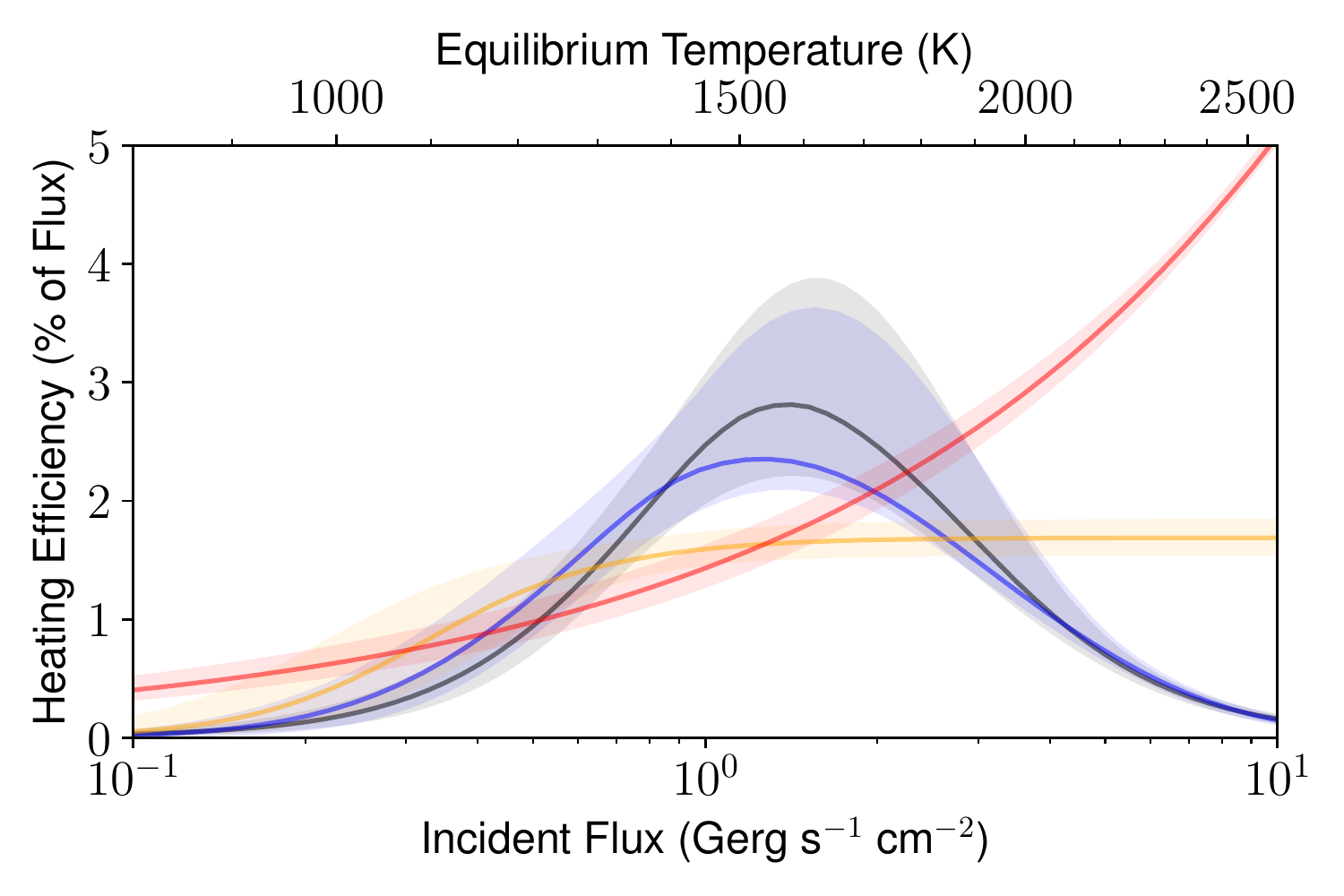}
    \caption{The posteriors of our statistical models of inflation power (as a percent of flux) against incident flux, with $1\sigma$ uncertainty bounds.  The red line is a power-law model, yellow is logistic, blue is Gaussian, and black is the GP.  The Gaussian model is strongly favored over the other parametric models by the DIC model selection criterion, and the GP strongly indicates a negative relationship at high flux.  This decrease in inflation efficiency at higher fluxes is important, because it matches predictions from the Ohmic dissipation mechanism of hot Jupiter inflation.}
    \label{fig:posteriors}
\end{figure}

The results for $\epsilon(F)$ are shown in Figure \ref{fig:posteriors}.  All functional forms yield similar results below about 0.5 \gerg, but differ significantly above this.  The GP model reaches a peak at around 1600 K and decreases towards zero with high statistical confidence, as shown by the uncertainty bounds.  At high fluxes, the uncertainty in heating power is roughly constant, and so declines as a fraction of flux.  Figure \ref{fig:massRadius} shows the predicted radius for a given mass of 5 Gyr old planets of average (posterior mean) composition and inflation power using the GP model.  The predictions align well with planets of similar mass and temperature.  The shape of $\epsilon(F)$ presented by the GP is corroborated by comparison of the DIC values. Of the parametric models, the Gaussian model is most favored, with a DIC of -1723.  The logistic model was next, at -1648, followed by the power-law model at -1641.  We interpret this to mean that $\epsilon$ decreases towards zero at high fluxes with high statistical significance, in agreement with our conclusions from the GP approach.  The DIC of the GP model is -1723, so there is no significant preference between it and the Gaussian model.  We present the Gaussian model since it takes a simple analytic form, as a percent of flux and with flux in units of \gerg:

\begin{equation}
    \epsilon = \left(2.37_{-.26}^{+1.3}\right)\mathrm{Exp}\left[
        -\frac{\left(\log(F)-\left(.14_{-.069}^{+.060}\right)\right)^2}{2\left(.37_{-.059}^{+.038}\right)^2}
    \right]
\end{equation}

Note that for planets whose interiors are in thermal equilibrium where $E_{in} = E_{out}$ and therefore $dR/dt = 0$ (which may happen quite early -- see Fig. \ref{fig:evolution}), the intrinsic temperature is directly related to $\epsilon$ as:
\begin{equation}
    T_{int} = \left(\frac{\epsilon F}{4 \sigma}\right)^\frac{1}{4} = \epsilon^\frac{1}{4} T_{\mathrm{eq}}
\end{equation}
where $\sigma$ is the Stefan-Boltzmann constant, and the conversion from flux to equilibrium temperature assumes an ideal black body with full heat redistribution.

\begin{figure}
    \centering
    \includegraphics[width=.47\textwidth]{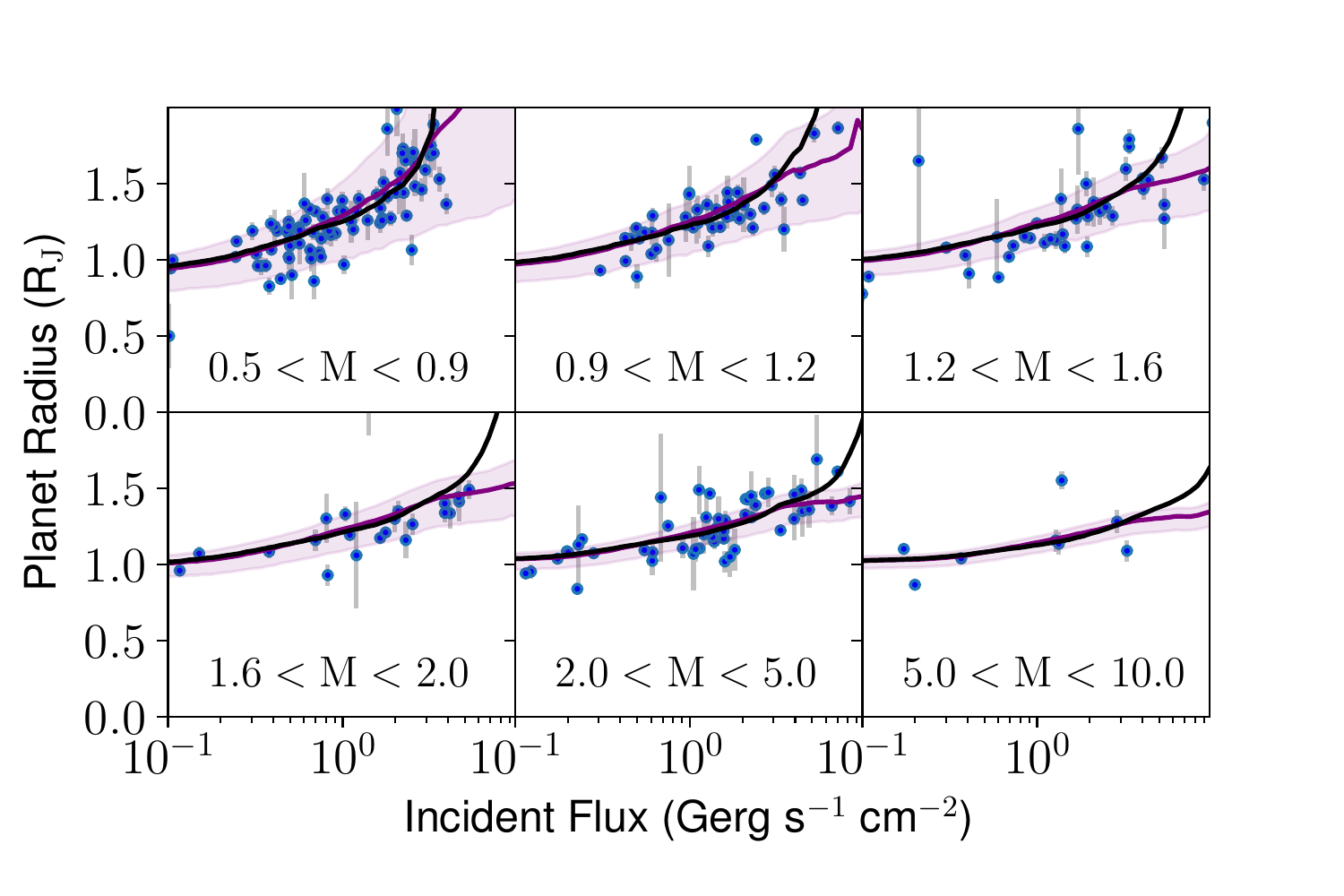}
    \caption{The radii of transiting giant planets against flux, divided into six mass bins.  The blue line and region are the Gaussian model's predicted radius and 1 $\sigma$ uncertainty bounds.  The black line is the prediction for the next best model, the logistic function.  The latter makes similar predictions but over-predicts radii of high flux planets, so the DIC favors the Gaussian model by a statistically significant margin.  This is more obvious looking directly at the residuals, which are shown in Fig. \ref{fig:resid}.}
    \label{fig:predictive}
\end{figure}

\begin{figure}
    \centering
    \includegraphics[width=.47\textwidth]{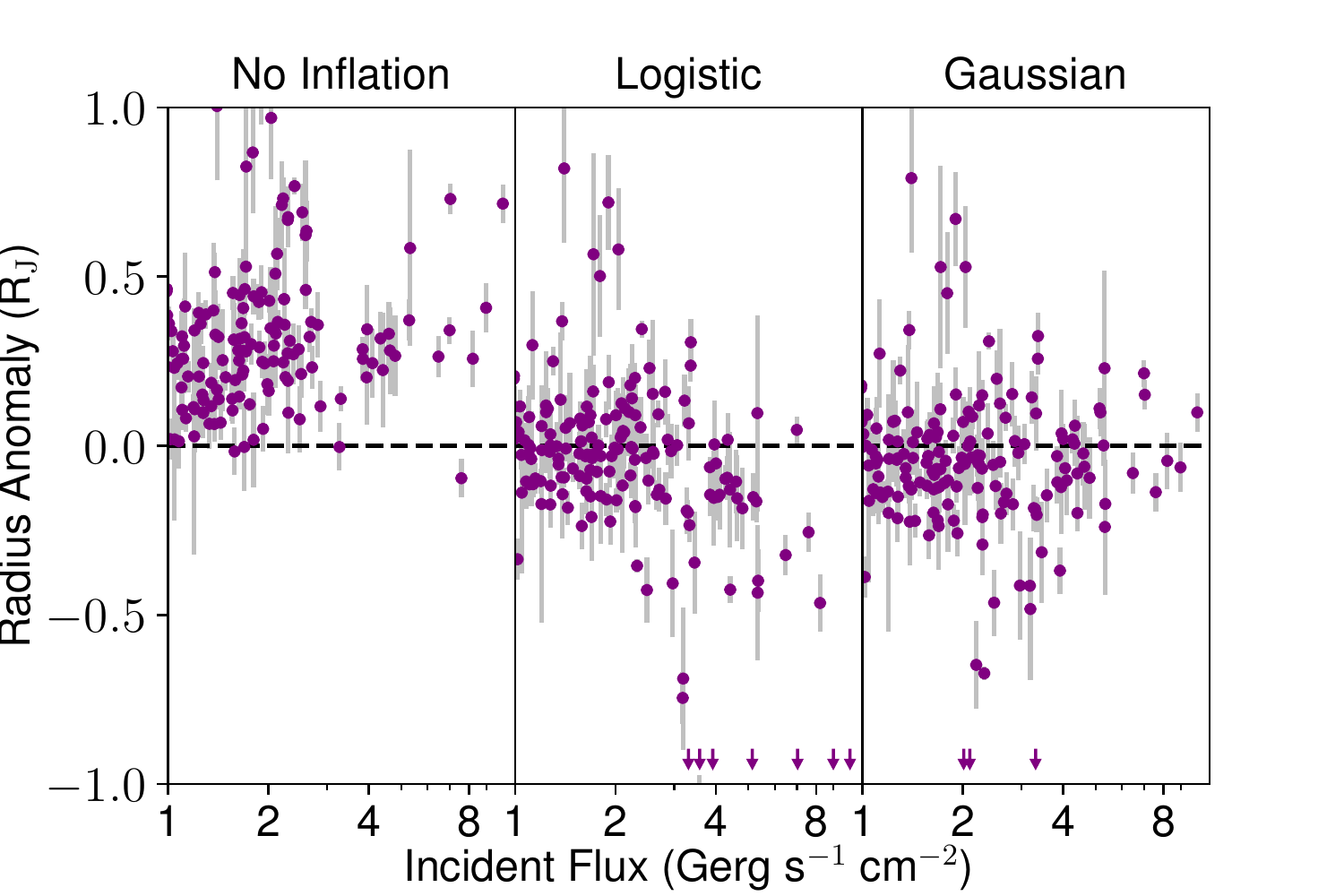}
    \caption{The difference between observed and predicted radius plotted against incident flux assuming typical composition planets under the cool giant model (no inflationary effect), the logistic model, and the Gaussian model (see Fig. \ref{fig:posteriors}).  Arrows show the handful of planets where the model exceeded the observed radius by more than 1 $\mathrm{R_J}$, which typically occurs only for very hot, very low mass planets whose radii are extremely sensitive to bulk metallicity.  Heavy-element abundance variations \cite{Thorngren2016} are sufficient to explain the scatter (see Fig. \ref{fig:predictive}). Error-bars depict observational error only.  The plot illustrates why our statistical tests prefer the Gaussian model over the logistic model: the logistic model consistently overestimates the radii of planets at high fluxes, while the Gaussian model does not.}
    \label{fig:resid}
\end{figure}

To visualize why the Gaussian model is preferred, we compute the posterior predictive radius distributions, and compare them to the radii of our observed planets.  Figure \ref{fig:predictive} compares these predictions for the favored GP model and the next-best logistic model to the observed radii as a function of incident flux, divided into six mass bins.  The models only diverge at high fluxes, about 2 \gerg.  Beyond this, the logistic model systematically overestimates the radii, and the GP does not.  To make this clear, Figure \ref{fig:resid} shows the residual to the expected radius (the radius anomaly) for high fluxes under a no inflation model, the logistic model, and the Gaussian model.  Here, the increasing bias of the logistic model for the ~30 planets at such high fluxes is apparent.  Even a flat $\epsilon$ at high flux predicts overly large planets, hence our conclusion that $\epsilon(F)$ must decline.

\begin{figure}
    \centering
    \includegraphics[width=.47\textwidth]{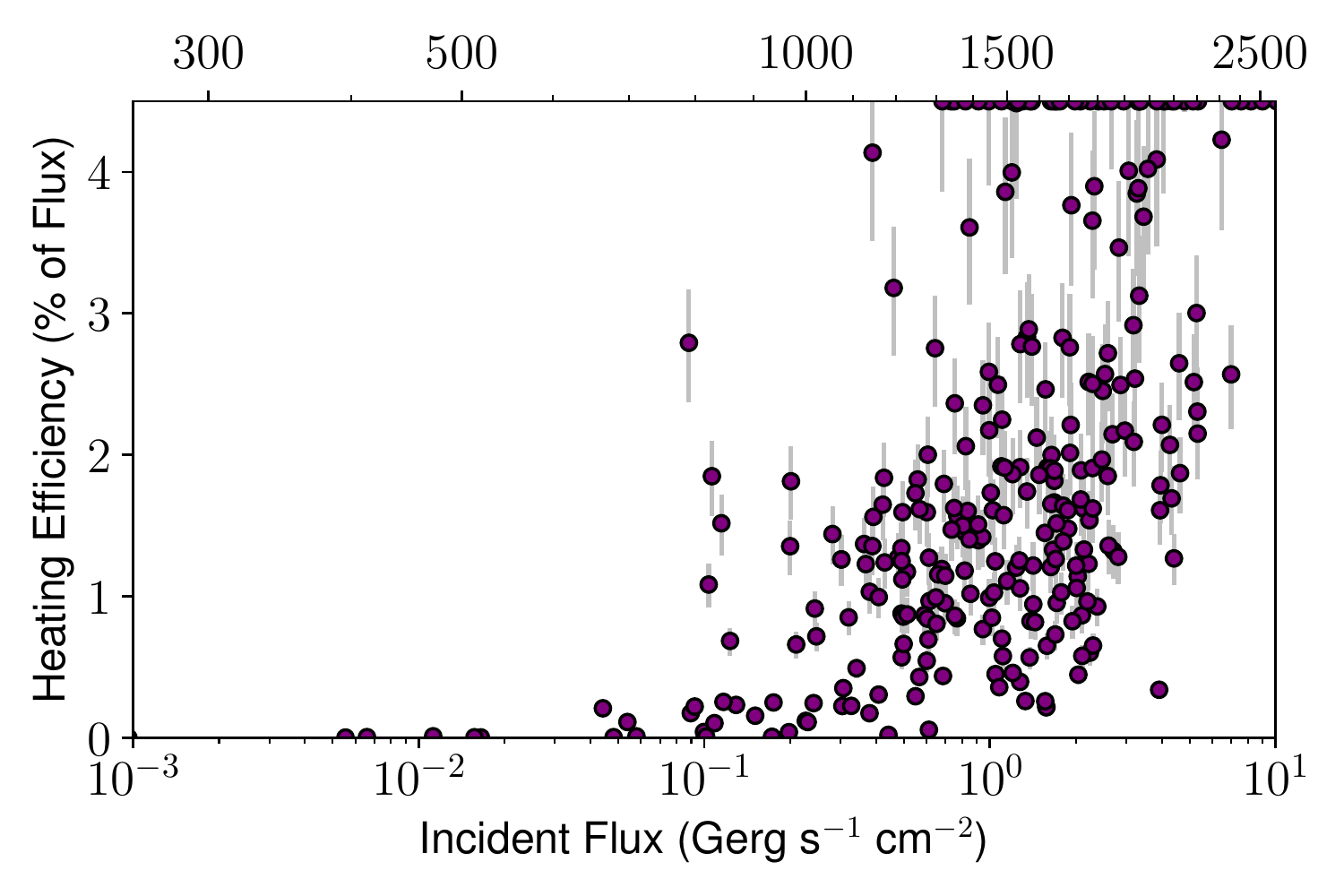}
    \caption{The posterior heating efficiencies for our sample planets as a function of flux, using the thermal tides scaling relationship from \cite{Socrates2013} but leaving a constant scaling factor as a fit parameter.  The match with observations was poor, as it forces $\epsilon$ to vary by orders of magnitude in ways not apparent in the planet radii.  The DIC was -1642, much lower than the GP or gaussian models, though this was likely affected by our constraint that $\epsilon < 4.5\%$.  A large fraction of the data (43\% or 122/281   ) exceeded this upper bound and was clipped down to 4.5\%.}
    \label{fig:tides}
\end{figure}

For our model of thermal tides \citep{Arras2009}, we examined the scaling relations for thermal tides from \cite{Socrates2013} (Eq. \ref{tidesModel}), and found this potential power source to much too strongly increase with flux to reproduce the observed radii.  The variance also appears overly high; for example, the scaling relations force $\epsilon$ to vary by more than an order of magnitude just in planets with fluxes between 0.8 and 1.2 \gerg.  As a result, we encountered considerable difficulty getting the model (see section \ref{S:statsModels}) to fit.  We were only able to fit a model by imposing the regularizing constraint that $\epsilon$ for any individual planet cannot exceed $4.5\%$, a level far above what is otherwise needed to explain the observed radii.  Under this requirement, we measure $\log_{10}(\epsilon_0)=-1.61 \pm .065$, Figure \ref{fig:tides} shows the the inferred heating efficiencies for the sample planets as a function of flux.  The MCMC was able to fit the bulk of the data by placing them in the $.5-3 \%$ range, but the scaling is far too extreme.  In explaining the bulk of the planets, a huge 43\% (122/281) of the data exceeded the upper bound.  Without the constraint, very few of the planets actually end up inflated; the range of coefficients to $\epsilon_0$ given by the scaling relation from \cite{Socrates2013} is simply too large.  As such, we conclude that the dominant source of inflation power in the observed population does not follow the thermal tides scaling relation.

\section{Discussion}
The Gaussian shape is significant because it exclusively matches predictions of hot Jupiter inflation from the Ohmic dissipation mechanism.  Under this model, magnetic interactions transfer energy from the atmosphere of a planet into its interior \citep{Batygin2010}.  The effect is initially increasing with greater atmospheric temperatures and therefore ionization, but at very high temperatures the magnetic drag on atmospheric winds \citep{Perna2010} inhibit the process \citep{Menou2012,Batygin2011}.  \cite{Batygin2011} predicts a scaling with equilibrium temperature as $\epsilon \propto (1500\rm{K}/T_{\rm{eq}})^4$.  Menou also derives scaling laws for this effect, estimating the peak $\epsilon$ to occur at ~1600 K, depending on the planetary magnetic field strength \citep{Menou2012}.  \cite{Ginzburg2016} supports this conclusion, estimating a peak $\epsilon$ to occur at ~1500 K, with power-law tails on either side.  Finally, MHD simulations in \cite{Rogers2014} find a peak at 1500-1600 K. Figure \ref{fig:posteriors} shows that the posteriors of our favored models match these predictions well.  If Ohmic dissipation is responsible for our observation, then our measured $\epsilon(F)$ is presumably the average over various planetary magnetic field strengths.

A noteworthy difficulty with identifying our results with the Ohmic dissipation model is the depth at which the anomalous heat is deposited.  Our model assumes that anomalous heating is efficiently conducted into the interior adiabat.  Ohmic heating, however, is generally believed to be deposited at pressures low enough that only a portion of the deposited energy is inducted into the adiabat and a delayed cooling effect is produced \citep{Spiegel2013, Wu2013, Komacek2017}.  Indeed, \cite{Rogers2014} do not see sufficient heating to explain the observed radii.  As well as differing from our modeling assumptions, this appears inconsistent with the results of \cite{Hartman2016}, who observe re-inflation of giants as their parent stars age and brighten over their main-sequence lifetime.  This effect would be prohibitively slow in the shallow deposition case \citep{Ginzburg2016}.  Thus if Ohmic heating is to explain our results, it must either violate these predictions or be modified by an additional effect which ushers the heat further into the planet.  The advection effects proposed by \cite{Tremblin2017} show that such effects are plausible and that there is still a great deal left to understand about atmospheric flows in hot Jupiters.

As the results of \cite{Tremblin2017} stand, our observations to not seem to support them as the sole cause of inflation.  They predict observable inflation occurring well below the observed 0.2 \gerg threshold, and do not appear to support a decrease in efficiency at high flux.  However, our results might align better if temperature-dependent wind speeds are considered within their model, which could slow flows both at especially low and high $T_\mathrm{eq}$.  Slower winds at high  $T_\mathrm{eq}$ would be a natural consequence of magnetic drag \citep{Perna2010}.  We view our results here as support for the idea that magnetic drag is quite important in the hottest atmospheres.

Other candidate inflation models do not match our results very well.  Tidal heating may introduce non-negligible energy into planet interiors, but cannot fully explain the anomalous radii \citep{Miller2009,Leconte2010}, and would not reproduce our relationship with flux.  The thermal tides mechanism \citep{Socrates2013} appears to predict more variation in $\epsilon$ than can plausibly exist (see Fig. \ref{fig:tides}).  Delayed cooling models propose that no anomalous heating occurs and that radii anomalies instead result from phenomena which prevent the escape of formation energy, such as enhanced atmospheric opacities \citep{Burrows2007} or inefficient heat transport in the interior \citep{Chabrier2007}.  This energy would otherwise rapidly radiate away.  The issue with these proposals is that they do not inherently depend on flux and cannot explain the results of \cite{Hartman2016}.  Furthermore, in the case of layered convection \citep[see][]{Leconte2012} resulting in delayed cooling \citep{Chabrier2007}, structure evolution simulations in \cite{Kurokawa2015} show that layered convection would not occur in young giants, and that even if layers are imposed, they would need to be implausibly thin ($1-1000$cm) to achieve the observed radii.

The situation for Saturn-mass planets (those excluded from our model) remains puzzling.  As described in Section \ref{subSaturns}, these exhibit a different relationship with flux than Jupiter mass planets (Fig. \ref{fig:massRadius}) and have been found less frequently in high flux orbits than their higher-mass analogs (Fig. \ref{fig:massflux}).  Inefficiency in the heating mechanism, perhaps by lower magnetic field strengths, could explain the former observation, but not the latter.  Furthermore, \cite{Pu2017} recently showed that Ohmic dissipation should occur in Neptunes, so we can reasonably expect that it would work on Saturns as well.  Some observational biases are doubtless present, but would likely not produce the effects seen.  Thus it seems possible that mass loss is occurring.  However, the exact mechanism would be unclear; for example, neither XUV driven mass-loss \citep{Yelle2004,Lopez2012} nor boil-off \citep{Owen2016} appear to significantly affect planets in this mass range.  As such, the cause of these observations is an open question.

There is still much work to be done in understanding hot Jupiter radius inflation.  A promising avenue are the case of ``reinflated'' hot Jupiters, which are planets whose radii may be increasing over time as their stars evolve off the main sequence and brighten \citep{Lopez2016}.  \cite{Grunblatt2017} have conducted promising observations of two potentially re-inflated planets around sub-giant stars.  Our posterior radius predictions are closer to their observations under the re-inflated case, but more planets will be needed to establish strong statistical significance.  Comparing the main-sequence reinflation results of \cite{Hartman2016} with structure models could reveal the timescale of re-inflation, which is closely related to the depth of energy deposition \citep{Komacek2017,Ginzburg2016}.  If re-inflation does indeed occur, delayed cooling models are ruled out.  Follow-up work of \cite{Tremblin2017} to determine how their results would be affected by temperature-dependant wind speeds would also be helpful.  Finally, further magnetohydrodynamic simulations are needed to properly understand heat flow in the outer layers of these planets.  Our results add to this picture by providing strong evidence of a heating efficiency drop at high temperatures and thereby pointing us towards the Ohmic dissipation model; they also suggest that 3-D atmospheric circulation models need to take magnetic fields into account.

\acknowledgments 
The authors thank Eric Lopez, Vivien Parmentier, Thad Komacek, and Ruth Murray-Clay for helpful discussions.  Funding for this work was provided by NASA XRP grant NNX16AB49G.
\bibliography{bibliography}

\begin{thebibliography}{}
\expandafter\ifx\csname natexlab\endcsname\relax\def\natexlab#1{#1}\fi

\bibitem[{{Akeson} {et~al.}(2013){Akeson}, {Chen}, {Ciardi}, {Crane}, {Good},
  {Harbut}, {Jackson}, {Kane}, {Laity}, {Leifer}, {Lynn}, {McElroy}, {Papin},
  {Plavchan}, {Ram{\'{\i}}rez}, {Rey}, {von Braun}, {Wittman}, {Abajian},
  {Ali}, {Beichman}, {Beekley}, {Berriman}, {Berukoff}, {Bryden}, {Chan},
  {Groom}, {Lau}, {Payne}, {Regelson}, {Saucedo}, {Schmitz}, {Stauffer},
  {Wyatt}, \& {Zhang}}]{Akeson2013}
{Akeson}, R.~L., {Chen}, X., {Ciardi}, D., {et~al.} 2013, \pasp, 125, 989

\bibitem[{{Arras} \& {Socrates}(2009)}]{Arras2009}
{Arras}, P., \& {Socrates}, A. 2009, ArXiv e-prints, arXiv:0901.0735

\bibitem[{{Baraffe} {et~al.}(2004){Baraffe}, {Selsis}, {Chabrier}, {Barman},
  {Allard}, {Hauschildt}, \& {Lammer}}]{Baraffe2004}
{Baraffe}, I., {Selsis}, F., {Chabrier}, G., {et~al.} 2004, \aap, 419, L13

\bibitem[{{Batygin} \& {Stevenson}(2010)}]{Batygin2010}
{Batygin}, K., \& {Stevenson}, D.~J. 2010, \apjl, 714, L238

\bibitem[{{Batygin} {et~al.}(2011){Batygin}, {Stevenson}, \&
  {Bodenheimer}}]{Batygin2011}
{Batygin}, K., {Stevenson}, D.~J., \& {Bodenheimer}, P.~H. 2011, \apj, 738, 1

\bibitem[{{Burrows} {et~al.}(2007){Burrows}, {Hubeny}, {Budaj}, \&
  {Hubbard}}]{Burrows2007}
{Burrows}, A., {Hubeny}, I., {Budaj}, J., \& {Hubbard}, W.~B. 2007, \apj, 661,
  502

\bibitem[{{Chabrier} \& {Baraffe}(2007)}]{Chabrier2007}
{Chabrier}, G., \& {Baraffe}, I. 2007, \apjl, 661, L81

\bibitem[{{Charbonneau} {et~al.}(2000){Charbonneau}, {Brown}, {Latham}, \&
  {Mayor}}]{Charbonneau2000}
{Charbonneau}, D., {Brown}, T.~M., {Latham}, D.~W., \& {Mayor}, M. 2000, \apjl,
  529, L45

\bibitem[{{Demory} \& {Seager}(2011)}]{Demory2011}
{Demory}, B.-O., \& {Seager}, S. 2011, \apjs, 197, 12

\bibitem[{Foreman-Mackey(2016)}]{foreman-mackey2016}
Foreman-Mackey, D. 2016, The Journal of Open Source Software, 24,
  doi:10.21105/joss.00024

\bibitem[{{Fortney} {et~al.}(2007){Fortney}, {Marley}, \&
  {Barnes}}]{Fortney2007}
{Fortney}, J.~J., {Marley}, M.~S., \& {Barnes}, J.~W. 2007, \apj, 659, 1661

\bibitem[{Gelman {et~al.}(2013)Gelman, Carlin, Stern, Dunson, Vehtari, \&
  Rubin}]{Gelman2013}
Gelman, A., Carlin, J., Stern, H., {et~al.} 2013, Bayesian Data Analysis, Third
  Edition, Chapman \& Hall/CRC Texts in Statistical Science (Taylor \& Francis)

\bibitem[{Gelman {et~al.}(2014)Gelman, Hwang, \& Vehtari}]{Gelman2014}
Gelman, A., Hwang, J., \& Vehtari, A. 2014, Statistics and Computing, 24, 997

\bibitem[{{Ginzburg} \& {Sari}(2016)}]{Ginzburg2016}
{Ginzburg}, S., \& {Sari}, R. 2016, \apj, 819, 116

\bibitem[{{Grunblatt} {et~al.}(2017){Grunblatt}, {Huber}, {Gaidos}, {Lopez},
  {Howard}, {Isaacson}, {Vanderburg}, {Nofi}, {Yu}, {North}, {Chaplin},
  {Foreman-Mackey}, {Petigura}, {Ansdell}, \& {Weiss}}]{Grunblatt2017}
{Grunblatt}, S.~K., {Huber}, D., {Gaidos}, E., {et~al.} 2017, ArXiv e-prints,
  arXiv:1706.05865

\bibitem[{{Guillot} \& {Showman}(2002)}]{Guillot2002}
{Guillot}, T., \& {Showman}, A.~P. 2002, \aap, 385, 156

\bibitem[{{Hartman} {et~al.}(2016){Hartman}, {Bakos}, {Bhatti}, {Penev},
  {Bieryla}, {Latham}, {Kov{\'a}cs}, {Torres}, {Csubry}, {de Val-Borro},
  {Buchhave}, {Kov{\'a}cs}, {Quinn}, {Howard}, {Isaacson}, {Fulton}, {Everett},
  {Esquerdo}, {B{\'e}ky}, {Szklenar}, {Falco}, {Santerne}, {Boisse},
  {H{\'e}brard}, {Burrows}, {L{\'a}z{\'a}r}, {Papp}, \&
  {S{\'a}ri}}]{Hartman2016}
{Hartman}, J.~D., {Bakos}, G.~{\'A}., {Bhatti}, W., {et~al.} 2016, \aj, 152,
  182

\bibitem[{Hastings(1970)}]{Hastings1970}
Hastings, W.~K. 1970, Biometrika, 57, 97

\bibitem[{{Henry} {et~al.}(2000){Henry}, {Marcy}, {Butler}, \&
  {Vogt}}]{Henry2000}
{Henry}, G.~W., {Marcy}, G.~W., {Butler}, R.~P., \& {Vogt}, S.~S. 2000, \apjl,
  529, L41

\bibitem[{{Hubbard} {et~al.}(2007){Hubbard}, {Hattori}, {Burrows}, {Hubeny}, \&
  {Sudarsky}}]{Hubbard2007}
{Hubbard}, W.~B., {Hattori}, M.~F., {Burrows}, A., {Hubeny}, I., \& {Sudarsky},
  D. 2007, \icarus, 187, 358

\bibitem[{Kass \& Raftery(1995)}]{Kass1995}
Kass, R.~E., \& Raftery, A.~E. 1995, Journal of the American Statistical
  Association, 90, 773

\bibitem[{{Komacek} \& {Youdin}(2017)}]{Komacek2017}
{Komacek}, T.~D., \& {Youdin}, A.~N. 2017, ArXiv e-prints, arXiv:1706.07605

\bibitem[{{Kurokawa} \& {Inutsuka}(2015)}]{Kurokawa2015}
{Kurokawa}, H., \& {Inutsuka}, S.-i. 2015, \apj, 815, 78

\bibitem[{{Laughlin} {et~al.}(2011){Laughlin}, {Crismani}, \&
  {Adams}}]{Laughlin2011}
{Laughlin}, G., {Crismani}, M., \& {Adams}, F.~C. 2011, \apjl, 729, L7

\bibitem[{{Leconte} \& {Chabrier}(2012)}]{Leconte2012}
{Leconte}, J., \& {Chabrier}, G. 2012, \aap, 540, A20

\bibitem[{{Leconte} {et~al.}(2010){Leconte}, {Chabrier}, {Baraffe}, \&
  {Levrard}}]{Leconte2010}
{Leconte}, J., {Chabrier}, G., {Baraffe}, I., \& {Levrard}, B. 2010, \aap, 516,
  A64

\bibitem[{{Lopez} \& {Fortney}(2016)}]{Lopez2016}
{Lopez}, E.~D., \& {Fortney}, J.~J. 2016, \apj, 818, 4

\bibitem[{{Lopez} {et~al.}(2012){Lopez}, {Fortney}, \& {Miller}}]{Lopez2012}
{Lopez}, E.~D., {Fortney}, J.~J., \& {Miller}, N. 2012, \apj, 761, 59

\bibitem[{{Menou}(2012)}]{Menou2012}
{Menou}, K. 2012, \apj, 745, 138

\bibitem[{{Miller} \& {Fortney}(2011)}]{Miller2011}
{Miller}, N., \& {Fortney}, J.~J. 2011, \apjl, 736, L29

\bibitem[{{Miller} {et~al.}(2009){Miller}, {Fortney}, \&
  {Jackson}}]{Miller2009}
{Miller}, N., {Fortney}, J.~J., \& {Jackson}, B. 2009, \apj, 702, 1413

\bibitem[{{Owen} \& {Wu}(2016)}]{Owen2016}
{Owen}, J.~E., \& {Wu}, Y. 2016, \apj, 817, 107

\bibitem[{{Perna} {et~al.}(2010){Perna}, {Menou}, \& {Rauscher}}]{Perna2010}
{Perna}, R., {Menou}, K., \& {Rauscher}, E. 2010, \apj, 719, 1421

\bibitem[{{Pu} \& {Valencia}(2017)}]{Pu2017}
{Pu}, B., \& {Valencia}, D. 2017, \apj, 846, 47

\bibitem[{Roberts \& Rosenthal(2004)}]{Roberts2004}
Roberts, G.~O., \& Rosenthal, J.~S. 2004, Probab. Surveys, 1, 20

\bibitem[{{Rogers} \& {Komacek}(2014)}]{Rogers2014}
{Rogers}, T.~M., \& {Komacek}, T.~D. 2014, \apj, 794, 132

\bibitem[{{Saumon} {et~al.}(1995){Saumon}, {Chabrier}, \& {van
  Horn}}]{Saumon1995}
{Saumon}, D., {Chabrier}, G., \& {van Horn}, H.~M. 1995, \apjs, 99, 713

\bibitem[{{Schneider} {et~al.}(2011){Schneider}, {Dedieu}, {Le Sidaner},
  {Savalle}, \& {Zolotukhin}}]{Schneider2011}
{Schneider}, J., {Dedieu}, C., {Le Sidaner}, P., {Savalle}, R., \&
  {Zolotukhin}, I. 2011, \aap, 532, A79

\bibitem[{{Silverman}(1986)}]{Silverman1986}
{Silverman}, B.~W. 1986, {Density estimation for statistics and data analysis}

\bibitem[{{Socrates}(2013)}]{Socrates2013}
{Socrates}, A. 2013, ArXiv e-prints, arXiv:1304.4121

\bibitem[{Spiegel \& Burrows(2013)}]{Spiegel2013}
Spiegel, D.~S., \& Burrows, A. 2013, The Astrophysical Journal, 772, 76

\bibitem[{Spiegelhalter {et~al.}(2002)Spiegelhalter, Best, Carlin, \& Van
  Der~Linde}]{Spiegelhalter2002}
Spiegelhalter, D.~J., Best, N.~G., Carlin, B.~P., \& Van Der~Linde, A. 2002,
  Journal of the Royal Statistical Society: Series B (Statistical Methodology),
  64, 583

\bibitem[{{Thompson}(1990)}]{Thompson1990}
{Thompson}, S.~L. 1990, {ANEOS---Analytic Equations of State for Shock Physics
  Codes, Sandia Natl. Lab. Doc. SAND89-2951}

\bibitem[{{Thorngren} {et~al.}(2016){Thorngren}, {Fortney}, {Murray-Clay}, \&
  {Lopez}}]{Thorngren2016}
{Thorngren}, D.~P., {Fortney}, J.~J., {Murray-Clay}, R.~A., \& {Lopez}, E.~D.
  2016, \apj, 831, 64

\bibitem[{{Tremblin} {et~al.}(2017){Tremblin}, {Chabrier}, {Mayne}, {Amundsen},
  {Baraffe}, {Debras}, {Drummond}, {Manners}, \& {Fromang}}]{Tremblin2017}
{Tremblin}, P., {Chabrier}, G., {Mayne}, N.~J., {et~al.} 2017, \apj, 841, 30

\bibitem[{{Van Der Walt} {et~al.}(2011){Van Der Walt}, {Colbert}, \&
  {Varoquaux}}]{Scipy}
{Van Der Walt}, S., {Colbert}, S.~C., \& {Varoquaux}, G. 2011, ArXiv e-prints,
  arXiv:1102.1523

\bibitem[{{Weiss} {et~al.}(2013){Weiss}, {Marcy}, {Rowe}, {Howard}, {Isaacson},
  {Fortney}, {Miller}, {Demory}, {Fischer}, {Adams}, {Dupree}, {Howell},
  {Kolbl}, {Johnson}, {Horch}, {Everett}, {Fabrycky}, \& {Seager}}]{Weiss2013}
{Weiss}, L.~M., {Marcy}, G.~W., {Rowe}, J.~F., {et~al.} 2013, \apj, 768, 14

\bibitem[{Wu \& Lithwick(2013)}]{Wu2013}
Wu, Y., \& Lithwick, Y. 2013, The Astrophysical Journal, 763, 13

\bibitem[{{Yelle}(2004)}]{Yelle2004}
{Yelle}, R.~V. 2004, \icarus, 170, 167

\end{thebibliography}
\end{document}